\documentclass[12pt]{article}  
\usepackage{graphicx}
\newcommand{\vv}{\vspace*{1.5ex}}

\newcommand{\no}{\noindent}
 \newcommand{\bc}{\begin{center}}
 \newcommand{\ec}{\end{center}}
                   \newcommand{\bfr}{\begin{flushright}}
                   \newcommand{\efr}{\end{flushright}}
   
     \newcommand{\be}{\begin{enumerate}}
     \newcommand{\ee}{\end{enumerate}}
        \newcommand{\bi}{\begin{itemize}}
        \newcommand{\ei}{\end{itemize}}
            \newcommand{\bd}{\begin{description}}
            \newcommand{\ed}{\end{description}}
                \newcommand{\beq}{\begin{equation}}
                \newcommand{\eeq}{\end{equation}}
                  \newcommand{\bea}{\begin{eqnarray}}
                  \newcommand{\eea}{\end{eqnarray}}

      \newcommand{\bfi}{\begin{figure}}
      \newcommand{\efi}{\end{figure}}
\newcommand{\bay}{\begin{array}{l}}
\newcommand{\eay}{\end{array}}
            
\newcommand{\KK}{\mbox{\boldmath $K$}}
        
\begin{document}   \vskip 1.5in

 \begin{center}
 {\Large {\sf   Theory of Sorption Hysteresis
                 in Nanoporous Solids:\\[.1mm] 
                 II. Molecular Condensation
}}    \\[7mm]
{\sc
Martin Z. Bazant\footnote{Associate Professor of Chemical Engineering and
Mathematics, Massachusetts Institute of Technology, Cambridge MA 02139.}
and
Zden\v ek P. Ba\v zant\footnote{ McCormick Institute Professor and
W.P. Murphy Professor of Civil Engineering and Materials Science,
Northwestern University, 2145 Sheridan Road, CEE/A135, Evanston, Illinois
60208; z-bazant@northwestern.edu (corresponding author).} }

\today

\end{center} \vskip 5mm   

\noindent {\bf Abstract:}\, {\sf Motivated by the puzzle of sorption
hysteresis in Portland cement concrete or cement paste, 
we develop in Part II of this study 
a general theory of vapor sorption and desorption from nanoporous solids,
which attributes hysteresis to hindered molecular condensation with
attractive lateral interactions. The classical mean-field theory of 
van der Waals is applied to predict the dependence of hysteresis on
temperature and pore size, using the regular solution model and gradient
energy of Cahn and Hilliard. A simple ``hierarchical wetting" model for
thin nanopores is developed to describe the case of strong wetting by the
first monolayer, followed by condensation of nanodroplets and nanobubbles
in the bulk. The model predicts a larger hysteresis
critical temperature and enhanced hysteresis for molecular condensation across nanopores at high vapor pressure than within  monolayers at low vapor pressure. For heterogeneous
pores, the theory predicts sorption/desorption sequences similar to those
seen in molecular dynamics simulations, where the interfacial energy (or
gradient penalty) at nanopore junctions acts as a free energy barrier for
snap-through instabilities. The model helps to quantitatively understand
recent experimental data for concrete or cement paste   
wetting and drying cycles and suggests new experiments at different
temperatures and humidity sweep rates.
 }

\subsection*{Introduction}

\begin{figure}
\begin{center}
(a)\includegraphics[width=3in]{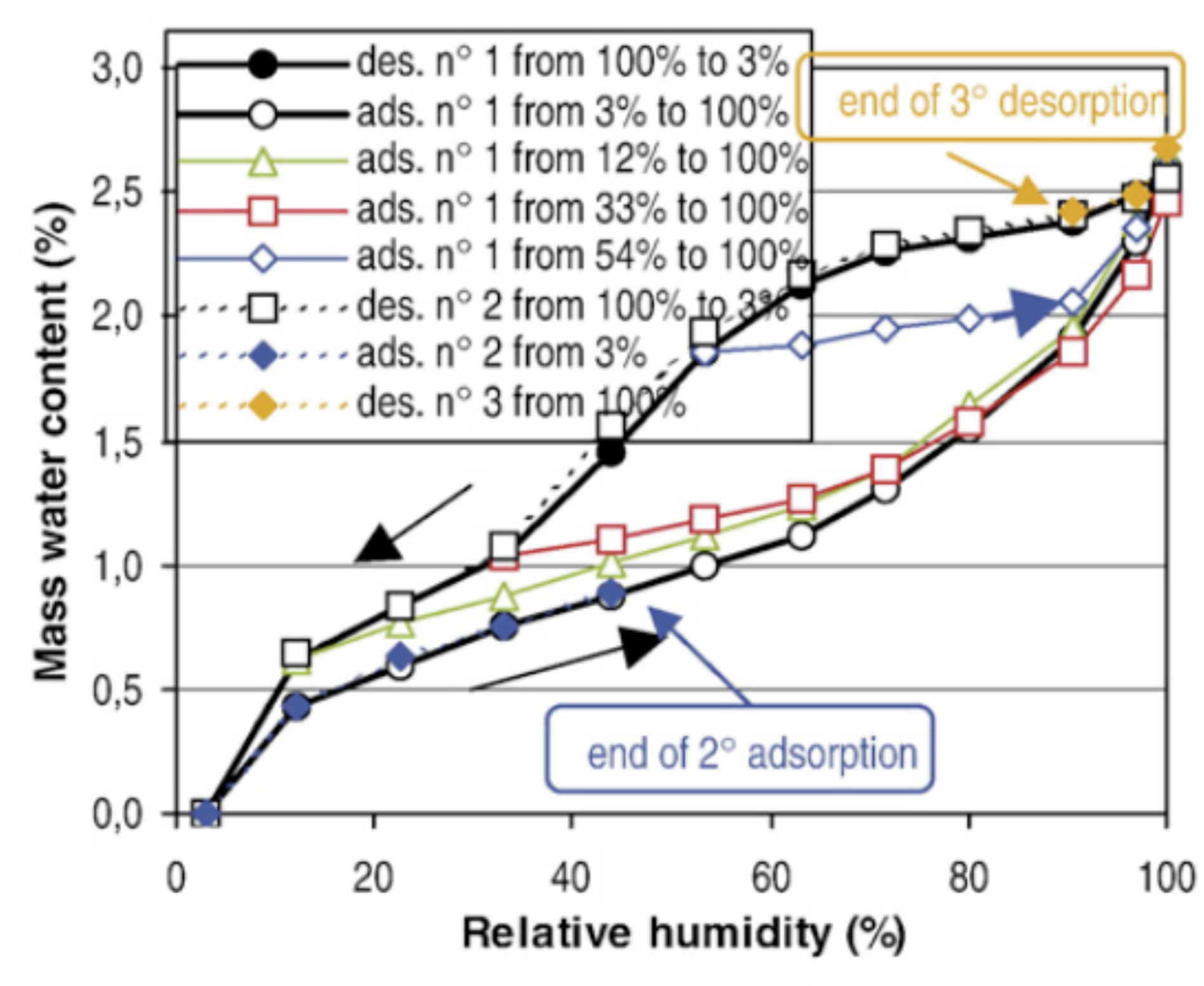}
(b)\includegraphics[width=3.2in]{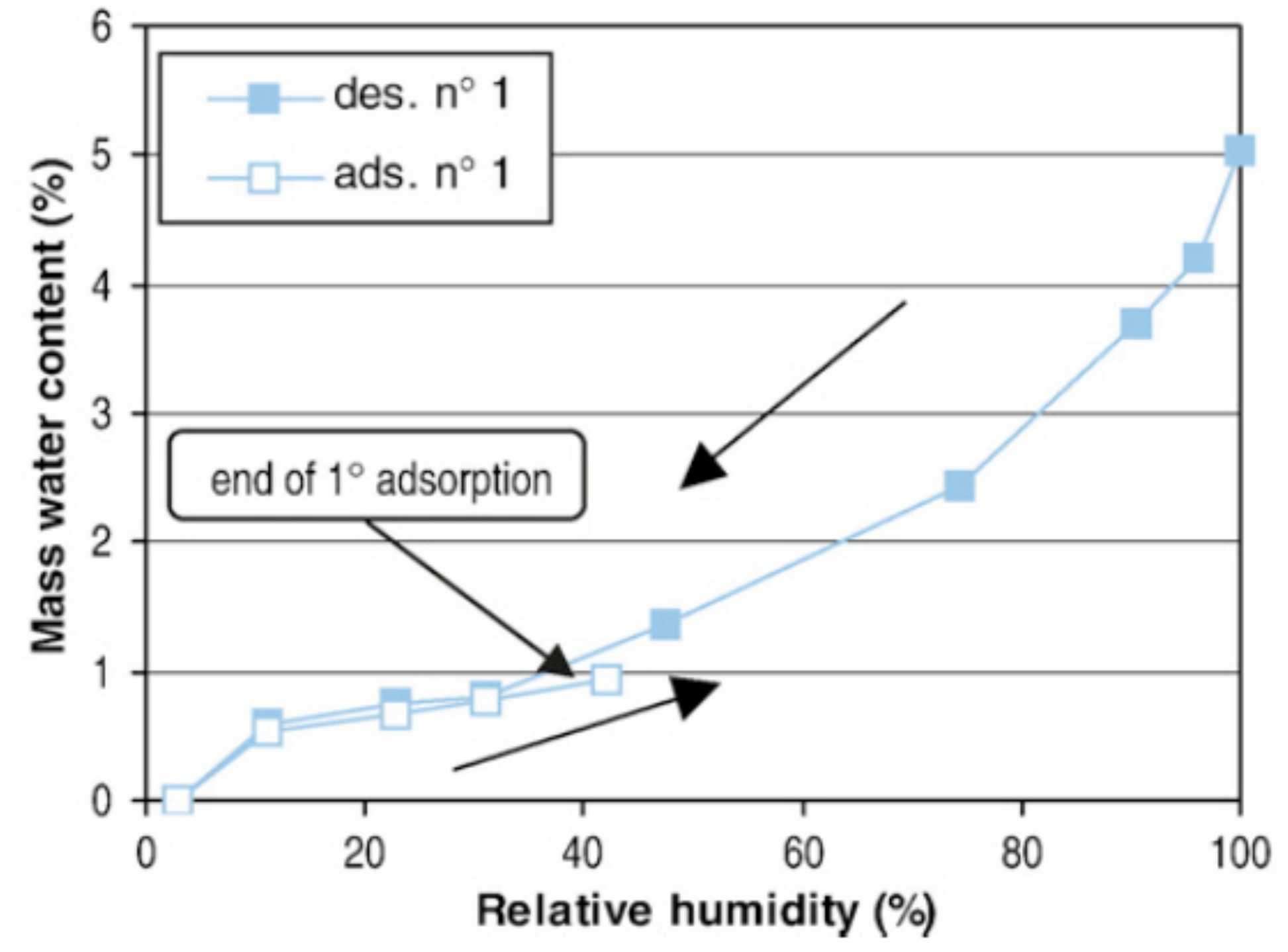}
 \caption{ Experimental data for isothermal water-vapor desorption and
sorption cycles in concrete at different temperatures. (a) Pronounced
hysteresis at $T = 23^\circ$C. (b) Suppressed hysteresis at $T =
44^\circ$C. [Reproduced from Fig. 2(e) and 2(g) of
Baroghel-Bouny~\cite{baroghel2007} with permission.] \label{fig:expt} }
\end{center}
\end{figure}

As introduced in Part I~\cite{part1}, a long-standing puzzle in the
thermodynamics of concrete or cement paste     
and other nanoporous solids is the pronounced hysteresis of the
sorption/desorption isotherm at low vapor pressure
~\cite{PowBro46,FelSer64,FelSer68,Rar-Jen95,Sch99,Jen-Sch08,AdoSet-02,EspFra06}.
Typical experimental data for wetting/drying  cycles in concrete is shown
in Fig~\ref{fig:expt}, and similar behavior can be observed (or expected)
in many other important situations, such as water sorption in dry soils or
wood, carbon sequestration in porous absorbents, and natural gas recovery
from nanoporous shales.  At low vapor pressures, well below the saturation
pressure, very little bulk liquid exists in the larger pores, and 
so the observed hysteresis cannot be attributed to the classical ``ink-bottle effect"
of capillarity from continuum fluid mechanics~\cite{Coh38,Bru43}. Moreover,
in nanopores, the Laplace tension 
of a continuous meniscus can easily exceed the tensile strength of the
liquid. So  
it is widely believed that adsorbate layers must be uniformly spread over
the entire internal surface area at low vapor pressure and unable to
coalesce into nonuniform patches or droplets.

This thinking underlies the ubiquitous method of determining the internal
surface area of porous media by fitting the sorption isotherm to the
Brunauer-Emmett-Teller (BET) equation of state~\cite{BET38}, which is
strictly valid only for a statistically homogeneous adsorbate on a flat
surface.  Since the BET isotherm is perfectly reversible, the internal
surface area can only be unambiguously inferred from one type of
measurement, either sorption or desorption, starting from a well defined
reproducible initial state.  Typically, the BET fit is made for sorption
starting from very low vapor pressure, assuming that the internal surface
is initially bare, but it is troubling to neglect the desorption data,
which would imply a different BET internal surface area, without a theory
to explain its origin. Moreover, if a mathematical theory could be
developed, then in principle one could extract more complete information
about the internal pore structure, such as the statistical distributions of
pore thickness or pore area, from the history dependence of sorption and
desorption.

If one insists on the validity of any reversible adsorption isotherm (not
only BET), then the only way to explain the observed hysteresis is to
invoke changes in the accessible internal surface area, e.g. due to
chemical transformations or structural
damage~\cite{FelSer68,Tho-Jen08,Rar-Jen95,EspFra06,Jen10}. As a result,
this picture of ``pore collapse" and subsequent reopening upon desorption
and sorption, respectively, is firmly entrenched, but it is noteworthy
that, sixty years after the first observations of sorption hysteresis in
concrete and cement paste, 
no mathematical theory has emerged to justify this assumption or make any
testable theoretical predictions. For example, it is not clear how the pore
collapse hypothesis could explain the strong dependence of sorption
hysteresis in concrete on temperature and chemical composition of the vapor
observed in recent experiments~\cite{baroghel2007}, as shown in Figure 1,
or how it could be reconciled with the measured shrinkage values. While
some adsorption-related structural changes surely occur in cement paste 
and concrete and other nanoporous solids, especially at high vapor
pressures, the repeatability of sorption hysteresis (after the first few
cycles) and the relatively small concomitant macroscopic deformations seem
inconsistent with the very drastic changes to the pore structure at the
nanoscale required by the pore-collapse hypothesis.

In this work, we show that, due to molecular discreteness, hysteresis is a
natural and unavoidable feature of sorption in nanoporous solids with {\it
fixed} pore geometries. In Part I~\cite{part1}, we showed that misfit
pressures 
due to discrete molecular forces around heterogeneities in the nanopore
geometry generally provide local energy barriers for the passage of the
adsorbate-vapor interface, consistent with evidence from molecular dynamics
simulations~\cite{Pel-Lev02,Coa-Pel07,Coa-Pel08,Coa-Pel08b,Coa-Pel09}. As
the thermodynamic driving force is increased by changing the vapor
pressure, the interface remains pinned at the heterogeneity until a sudden
``snap-through instability" occurs, analogous to snap-through buckling of a
flat arch. This theory is also reminiscent of the Peierls-Nabarro model of
dislocation motion in crystals, where the misfit strain energy due to
discrete molecular forces in the dislocation core provides the crucial
resistance to dislocation motion, which cannot be predicted by continuum
elasticity~\cite{hirth}. A common feature of both theories is the
assumption of a layered solid-like material undergoing sudden, localized
rearrangements in response to a driving force that overcome the effective
``lattice resistance".  An important difference between nanopore sorption
and dislocation motion, however, is that there is no reference crystal
structure or long-range order in the adsorbate, 
and so much more dramatic molecular rearrangements, such as wetting phase
transformations, are possible.  Predicting such phase transformations and
their dependence on temperature requires a more detailed molecular model.

Here, in Part II, we consider sorption from the general perspective of
statistical thermodynamics and develop a simple mathematical theory that
connects hysteresis to inter-molecular forces. The model is quantitatively
consistent with the concrete sorption
data in Figure 1 and suggests new directions for experiments and
simulations to further develop the theory. The key insight is that sorption
hysteresis is possible at sufficiently low temperature in any fixed surface
geometry, as long as the adsorbed molecules have a short-range attraction.
Although weaker than the orthogonal forces that bind the adsorbate to the
surface, such attractive lateral forces within the adsorbate itself promote
condensation into stable high density patches below a critical temperature,
regardless of the pore geometry. This phenomenon can be inhibited by
geometrical or chemical heterogeneities on the surface, but molecular
condensation can also occur in homogeneous pores or on flat surfaces, as
the metastable homogeneous adsorbate phase separates into stable
low-density and high-density phases within the porous structure.

Before we begin, let us explain our choice of terminology. The term
``capillary condensation" has been used to describe wetting/de-wetting
transitions on surfaces, which comprise a well-studied class of phase
separation phenomena in confined systems~\cite{gelb1999}. We avoid the use
of this term because in many fields, such as cement  
and concrete research (which motivates our work), the term ``capillary
water" refers to liquid water at high vapor pressure in large ($> 1 \mu$m)
pores, which can be modeled by continuum fluid mechanics with constant
gas/liquid surface tension. Here, we suggest the term ``molecular
condensation" to refer to the phase separation of discrete adsorbed
molecules in nanopores at low vapor pressure, which requires a statistical
mechanical treatment.

\subsection*{ Mean-Field Theory of Molecular Condensation  }

 {\bf Capillarity at the Molecular Scale:}
The macroscopic continuum theory of capillarity cannot be applied to very
thin adsorbate layers, whose individual molecules interact strongly with
the surface -- and each
other~\cite{israelachvilli,rowlinson,degennes1985,gelb1999}. The density of
an adsorbate is generally heterogeneous and lies between that of the bulk
liquid and vapor phases, due to attractive forces with the surface which
stabilize individual adatoms and (upon contact with a second surface) give
rise to disjoining pressure. These ``orthogonal forces" allow adsorbed
molecules to be distributed over a surface without maintaining close
lateral contacts. In a nanoscale pore, the total energy of the missing
lateral bonds would be grossly overestimated, if they were approximated by
sharp, continuous surfaces using the bulk surface tension and the nanoscale
curvature. Instead, one must develop a theory that takes ``lateral forces"
between discrete adsorbate molecules explicitly into account.

\begin{figure*}
\begin{center}
\includegraphics[width=5in]{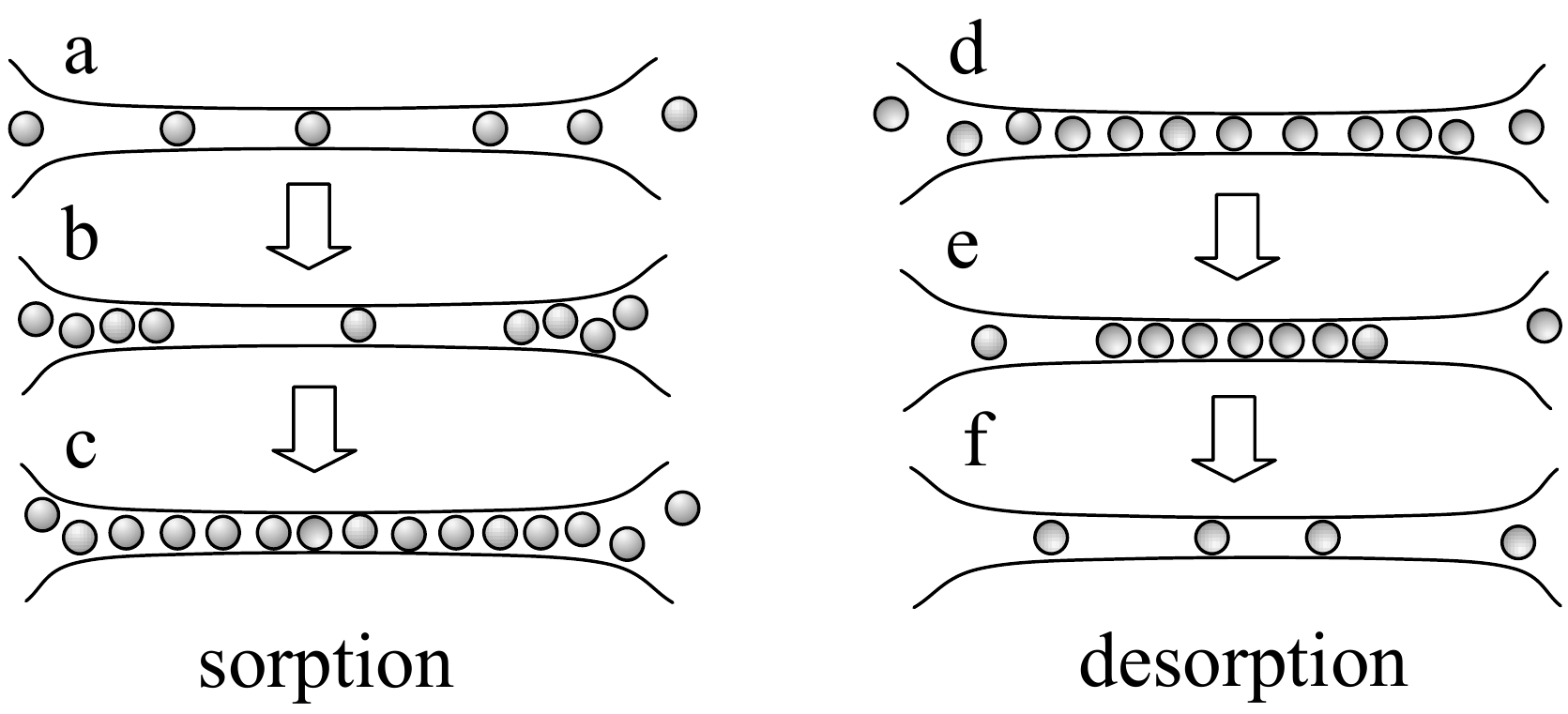}
\end{center}
\caption{  Molecular condensation in a straight, monolayer-thick pore
during sorption (left) and desorption (right) from the vapor at low temperature. 
Attractive lateral forces lead to the spontaneous separation of high-density and
low-density adsorbate phases from metastable homogeneous phases. Analogous
thermodynamic instabilities of the adsorbate distribution would also occur
in thicker pores or flat surfaces, only across a different range of
relative humidities, depending on the free energies of adsorption and
lateral interaction. In non-uniform pores, the snap-through instability is
another manifestation of this general phenomenon, driven by attractive
lateral (or inclined) forces in the adsorbate. It naturally leads to
sorption hysteresis without invoking any changes in pore structure.
\label{fig:coal} }
\end{figure*}

The theory of snap-through instabilities in nonuniform pore geometries from
Part I is an example of such an approach, but the effect of attractive
lateral forces is much more general and can lead to sorption hysteresis
even in perfectly uniform geometries.  The basic idea is already
illustrated by the simplest case of monolayer adsorption on a flat bare
surface or monolayer-thick pore, as shown in Figure ~\ref{fig:coal}. (The
latter problem is equivalent to lithium insertion and extraction in a
crystalline nanoparticle in a Li-ion
battery~\cite{singh2008,burch2009,bai2011}, and we apply similar concepts
and models for adsorption dynamics~\cite{MZB_notes}.)  As humidity
increases during sorption (left), the dilute homogeneous adsorbate $a$
becomes thermodynamically unstable and separates $b$ into locally stable
low-density and high-density phases, which quickly grow and merge into a
stable homogeneous adsorbate at high density $c$. As humidity then
decreases during desorption (right), the homogeneous phase $d$ destabilizes
and coalesces to form the stable, dense phase $e$, which quickly shrinks
and leaves behind a stable, homogeneous low-density adsorbate $f$. The
sketches in Figure ~\ref{fig:coal} assume nucleation of the second phase at
the pore openings, although other phase-separation pathways are possible.

The key point is that molecular condensation, or separation into
low-density and high-density adsorbate phases, is history dependent and
occurs by triggering the sudden instability of a metastable state. The
specific pore geometry is largely irrelevant. Spontaneous phase separation
of the adsorbate is mathematically analogous to the snap-through
instability of shells and arches, 
but the physical interpretation in terms of buckling failure may not always
apply. More importantly, in order to predict the effect of temperature on
sorption hysteresis, one must go beyond mechanical analogs and consider the
statistical thermodynamics of adsorption.

 {\bf Thermodynamics of Adsorption with Lateral Forces: }
Consider a nanopore or surface film, whose state is described by a
dimensionless filling $\Theta = \Gamma_w/\Gamma_1$, which may depend on
lateral position $x$. The Gibbs free energy per surface site $g$ can be
expressed for a homogeneous adsorbate as follows:
 \begin{equation}
  g_{hom}(\Theta,x) = g_{mix}(\Theta,x) - \Delta g_a(x) \, \Theta    \label{eq:ghom}
 \end{equation}
where $g_{mix}$ is the free energy of mixing in the pore volume $V_s(x)$
associated with surface site $x$, which contains an expected number
$\Theta(x)$ of adsorbate molecules, whose free energy change (per molecule)
due to adsorption from the vapor phase is
 \begin{equation}
  \Delta g_a = k_B T \ln c_0 + \Delta q_a.
 \end{equation}
The first term is a reference entropy, expressed in terms of a
dimensionless concentration $c_0$ (as in BET theory) and Boltzman's
constant $k_B$, and the second term, $\Delta q_a$, is the latent heat of
adsorption minus that of liquefaction per site. 
To focus on molecular effects, here in Part II we define energies per
particle, rather than per mole as in Part I ($\frac{\Delta Q_a }{RT}
=\frac{\Delta q_a}{k_BT}$), and we drop the overbars for ease of notation.
Lateral interactions among adsorbate molecules are included $g_{mix}$, e.g.
according to the regular solution model of the next section. Orthogonal
surface-adsorbate forces are treated separately via $\Delta g_a$. In the
case of pairwise interactions, the enthalpic contribution can be expressed
as
 \begin{equation}
  \Delta q_a = \int_{V_s} d\vec{r} \int_S d\vec{r}_s \;
  \Phi^s(|\vec{r}-\vec{r}_s)\,  p_s(\vec{r}|\vec{r}_s)
 \end{equation}
where the $\Phi^s(r)$ is the pair potential between adsorbate and surface
molecules, $p_s(\vec{r}|\vec{r}_s)$ is the conditional probability density
of finding a surface molecule at $\vec{r}_s$ anywhere in the solid volume
$S$ given an adsorbate molecule at $\vec{r}$ in the pore volume $V_s(x)$
associated with site $x$. In an isotropic bulk liquid, the pair correlation
function $g(r)$ is defined by $p_s(\vec{r}|\vec{r}^\prime)=4\pi r^2 g(r)$
where $r=|\vec{r}-\vec{r}^\prime|$, but here the pore surface breaks
symmetry.

\begin{figure*}
\begin{center}
\vskip -1in
\includegraphics[width=6.5in]{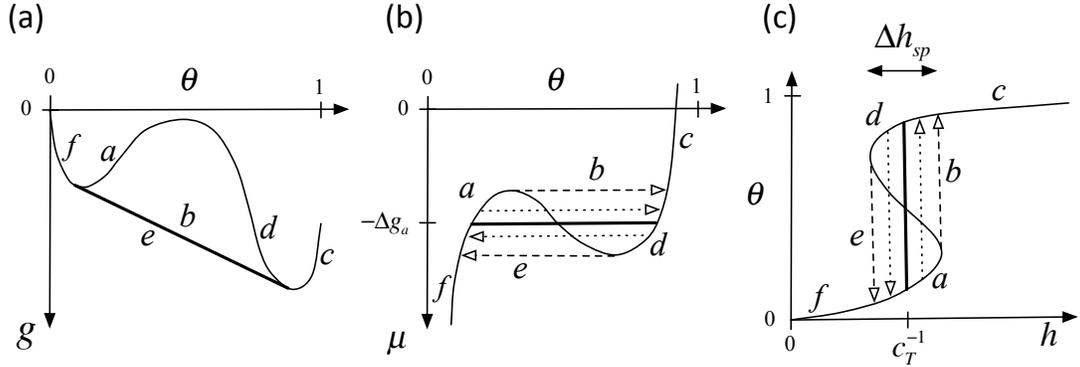}
\end{center}
 \vskip -1.8in
\caption{ Effects of lateral forces on adsorption. (a) Gibbs free energy
per site $g$ versus dimensionless filling $\Theta$ of adsorbate. The
homogeneous free energy $g_{hom}$ (thin solid line) is made convex by a
common tangent construction (thick solid line), which corresponds to phase
separation into high and low density regions (neglecting interphasial
tension).  (b) Chemical potential $\mu$ per site for homogeneous (thin
solid) and phase separated (thick solid) states. (c) The corresponing
filling fraction versus relative humidity $h$ for quasi-equilibrium between
the adsorbate and vapor. Hysteresis during adsorption ($a\to b \to c$ in
Fig. 10) or desorption ($d\to e \to f$) results from the delay in phase
separation due to either nucleation (dotted lines) or spinodal
decomposition (dashed lines). \label{fig:hyst} }
\end{figure*}

Due to attractive lateral interactions, at sufficiently low temperature the
homogeneous free energy of mixing $g_{mix}(\Theta)$ becomes non-convex and
leads to at least two local minima in the total free energy, corresponding
to stable high-density (liquid-like) and low-density (vapor-like) adsorbate
phases on the surface, as shown in Fig. ~\ref{fig:hyst}(a). A common
tangent construction connecting the two local minima, which restores
convexity across the ``miscibility gap", provides the mean free energy of a
phase separated system consisting of appropriate proportions of the
immiscible endpoint phases (neglecting interphasial tension, discussed
below). Phase  separation is illustrated by sketches in Figure
~\ref{fig:coal}, whose labels correspond to the letters in Figure
~\ref{fig:hyst}. During sorption, the homogeneous adsorbate passes the
low-density free energy minimum $a$, destabilizes and drops down to the
common tangent upon phase separation $b$, and becomes homogeneous again
after passing the high-density free energy minimum $c$. A similar
free-energy path is followed in reverse during desorption, only the
free-energy overshoot in the metastable homogeneous adsorbate occurs at
high density rather than low density. This is the fundamental source of
hysteresis.

The connection with sorption hysteresis becomes more clear from the
``diffusional" chemical potential of the homogeneous adsorbate,
 \begin{equation}     \label{eq:muhom}
  \mu_{hom} = \frac{dg_{hom}}{d\Theta} = \frac{dg_{mix}}{d\Theta} - \Delta g_a
 \end{equation}
which is the net free energy change to add a molecule (and remove any
vacant site). As sketched in Fig. \ref{fig:hyst}(b), a non-convex free
energy corresponds to a non-monotonic chemical potential versus
composition.  In equilibrium, the chemical potential of an adatom equals
that of a vapor molecule,
 \begin{equation}
  \mu_v = k_B T \ln h
 \end{equation}
where we set the zero of chemical potential in the saturated vapor phase
($h=1$). Setting $\mu_{hom} = \mu_v$ yields the equilibrium sorption curve
($\Theta$ vs $h$) for homogeneous filling of the nanopore, shown in Fig.
\ref{fig:hyst}(c). The non-convex free energy is seen to correspond to a
non-invertible sorption curve with three degenerate filling fractions over
the ``spinodal range" of humidities, $\Delta h_{sp}$. Over the
corresponding spinodal gap of filling fractions, where the free energy
loses convexity ($\frac{d^2 g_{hom}}{d\Theta^2} < 0$) and the chemical
potential decreases with concentration ($\frac{d\mu_{hom}}{d\Theta} < 0$),
the homogeneous adsorbate is linearly unstable with respect 
to the growth of infinitesimal perturbations of the concentration profile
(spinodal decomposition). This leads to sorption hysteresis with varying
humidity, as represented by the dashed lines in Figure ~\ref{fig:hyst}.

It is possible for phase separation to occur for any metastable composition
within the miscibility gap, but outside the spinodal gap, a sufficiently
large critical nucleus of the second phase is required. In the typical case
of heterogeneous nucleation, phase separation is triggered at nanopore
defects or boundaries, as sketched in Figure ~\ref{fig:coal}. If nucleation
occurs before spinodal decomposition, then phase separation occurs with
less overshoot of the chemical potential plateau and smaller sorption
hysteresis, as denoted by the dotted lines in Figure ~\ref{fig:hyst} (b)
and (c), respectively. For sufficiently slow humidity variations, the
nanopore should be able to reversibly follow the convex equilibrium free
energy surface without any hysteresis, but depending on experimental
conditions, the required nucleation and growth may not have enough time to
occur. In the absence of nucleation, the spinodal humidity range $\Delta
h_{sp}$ provides a convenient upper bound on the equilibrium sorption
hysteresis, and       
so we now proceed to calculate it using a simple model.

\begin{figure}
\begin{center}
(a) \includegraphics[width=5in]{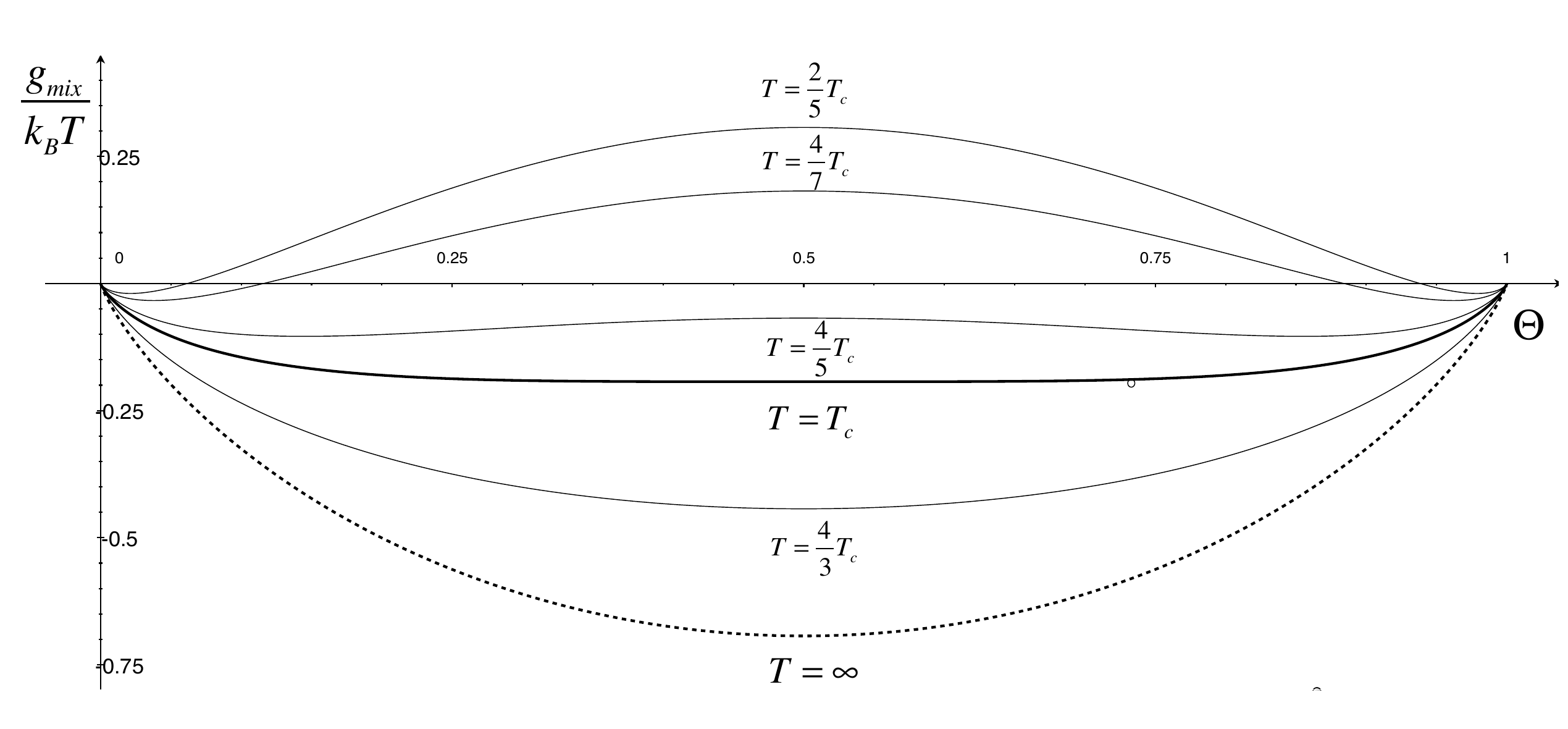}\\
\vspace{0.2in}
(b)\includegraphics[width=5in]{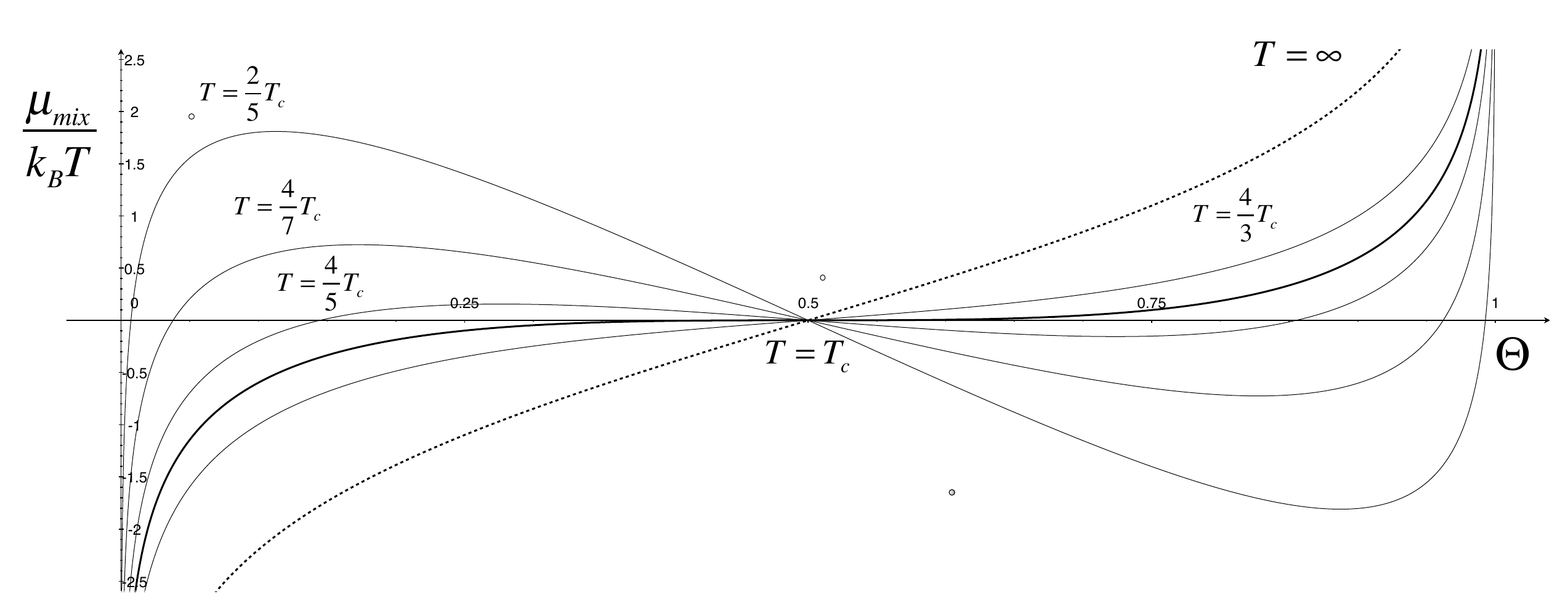}\\
\vspace{0.2in}
(c)\includegraphics[width=5in]{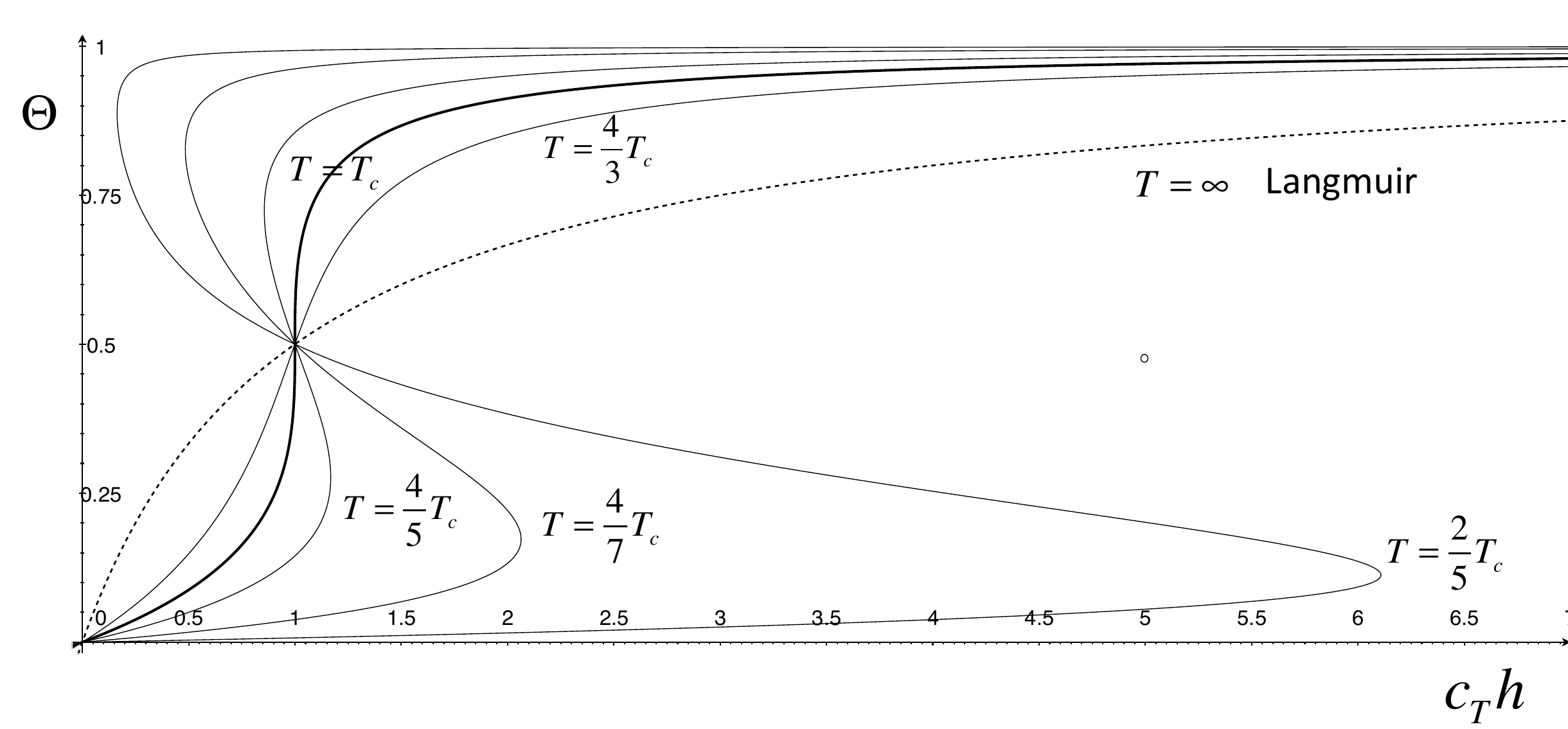}
\end{center}
\caption{ Thermodynamics of condensation in the regular solution model for
an adsorbed monolayer with attractive intermolecular forces ($\omega>0$).
(a) Free energy of mixing versus filling fraction. (b) Homogeneous chemical
potential (shifted by the adsorption free energy) versus filling fraction.
(c) Filling fraction versus humidity.  Below the critical temperature
$T_c=\frac{\omega}{2k_B}$, enthalpy dominates entropy; the free energy is
non-convex; the chemical potential is non-monotonic; and the adsorption
isotherms are multi-valued, leading to hysteresis.  \label{fig:regsol} }
\end{figure}

 {\bf Regular Solution Model:}
The simplest mean-field model of adsorption with lateral forces is the
regular solution model for a binary
mixture~\cite{guggenheim,CahHil58,baluffi}, whose free energy of mixing,
 \begin{equation}
  g_{mix} = k_B T \left[ \Theta \ln \Theta + (1-\Theta) \ln(1-\Theta)\right]
  + \omega \Theta (1-\Theta)
 \end{equation}
comes from the continuum limit of a lattice gas of filled and empty sites.
The first two terms represent the configurational entropy of particles and
holes in the lattice, and the last term represents the enthalpy of mixing,
expressed as a particle-hole interaction.  The lattice gas could represent
individual adsorbate molecules in a monolayer, either on a free surface or
in a flat one-molecule-thick nanopore as in Fig. ~\ref{fig:coal}. As
discussed below, the    
same model could also provide a first approximation of
hindered multilayer adsorption, where the particles and holes represent
coarsened  molecular droplets and bubbles spanning the interior of a
nanopore. Therefore, we will proceed to analyze sorption hysteresis in
general terms without yet referring a specific pore geometry.

Lateral adsorbate-adsorbate forces are captured by the regular solution
parameter, $\omega$, equal to the mean energy of pairwise attraction
between adsorbed molecules,
 \begin{equation}
  \omega= \int_{V_s} d\vec{r} \int_P d\vec{r}^\prime \; \frac{1}{2}
  \Phi(|\vec{r}-\vec{r}^\prime|)\,  p(\vec{r}|\vec{r}^\prime)
 \end{equation}
where $\Phi(r)$ is the pair potential between adsorbate molecules,
$p(\vec{r}|\vec{r}^\prime)$ is the conditional probability density of
finding a molecule at $\vec{r}^\prime$ anywhere in the pore volume $P$
given a molecule at $\vec{r}$ in the site volume $V_s$.  (The factor
$\frac{1}{2}$ avoids double counting pair interactions.) Note that $\Delta
g_a$ and $\omega$ depend on position in a heterogeneous pore, whose
geometry or surface chemistry varies with position.

As shown in Figure ~\ref{fig:regsol}(a), the homogeneous free energy of
mixing reflects a competition between entropy, which favors mixing ($\Theta
= \frac{1}{2}$) and enthalpy, which favors de-mixing or phase separation
($\Theta=0,1$). At high temperature, entropy dominates, and the free energy
is convex with a minimum at $\Theta = \frac{1}{2}$. Below a critical
temperature,
 \begin{equation}   \label{eq:Tc}
  T_c = \frac{\omega}{2k_B}
 \end{equation}
enthalpy (due to attractive lateral forces) begins to dominate entropy, and
there is a pitchfork bifurcation (in the mathematical sense), leading to
two local minima of the free energy density, corresponding to stable
high-density and low-density phases. The miscibility gap is the range of
metastable homogeneous compositions, bounded by the circles in
Fig.~\ref{fig:regsol}(a). The homogeneous chemical potential is given by
 \begin{equation}
  \mu_{hom} = k_B T \ln \frac{\Theta}{1-\Theta} + \omega(1- 2\Theta) - \Delta g_a
 \end{equation}
which is plotted for different temperatures in Figure ~\ref{fig:regsol}(b)
to illustrate the onset of non-monotonic behavior for $T < T_c$. The
spinodal gap is the range of unstable compositions, bounded by circles in
Fig.~\ref{fig:regsol}(b).

The corresponding adsorption isotherm, obtained by setting $\mu_{hom} =
\mu_v$, is given by
 \begin{equation}
  c_T h  = \left( \frac{\Theta_{hom}}{1-\Theta_{hom}} \right)\,
  \exp\left( \frac{\omega(1-2\Theta_{hom})}{k_B T} \right), \ \mbox{ where } \
  c_T = c_0 \exp\left(\frac{\Delta q_a}{k_B T}\right)
 \end{equation}
and plotted in Figure ~\ref{fig:regsol}(c). At high temperature, we recover
the classical Langmuir isotherm without lateral interactions,
 \begin{equation}
  \Theta_{hom} \sim \frac{c_T h}{ 1 + c_T h} \ \mbox{ for } \ T \gg T_c .
 \end{equation}
Below the critical point, $T < T_c$, the modified Langmuir isotherm with
lateral interactions becomes non-monotonic, and the sorption curve exhibits
hysteresis.

\begin{figure}[t]
\begin{center}
\includegraphics[width=5in]{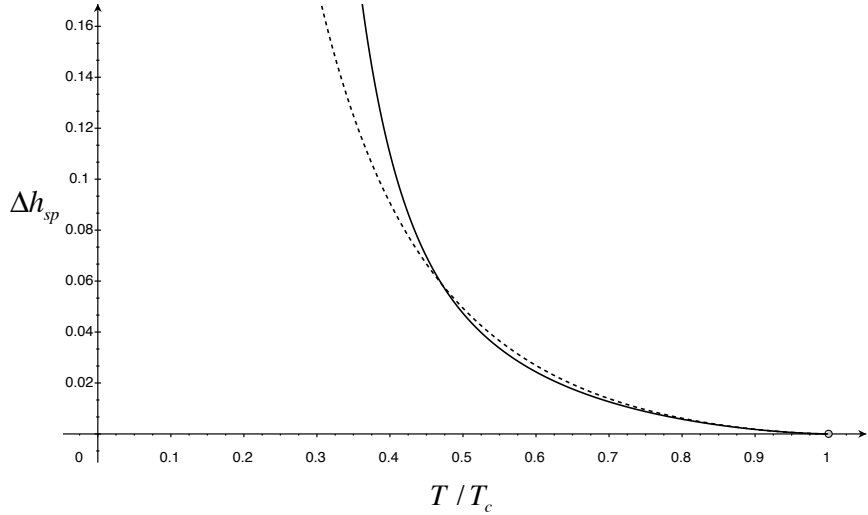}
\end{center}
 \vskip -0.7in
\caption{Simple analytical prediction of the temperature dependence of
sorption hysteresis, based on the regular solution model for lateral
interactions in the adsorbate. The change in relative humidity across the
spinodal range $\Delta h_{sp}$ is plotted against reduced temperature
$T/T_c$, where we set $c_T = 54$ for monolayer water adsorption in cement 
paste or concrete. We also neglect the weak temperature dependence of $c_T$
since adsorption forces are much stronger than lateral forces ($\Delta q_a
\gg 2 k_B T_c=\omega$). The exact solution (\ref{eq:dh1})-(\ref{eq:dh2})
(solid curve) is well approximated over this range by the asymptotic power
law at the critical point (\ref{eq:power}) (dashed curve). At higher
temperatures, $T > T_c$, molecular condensation is thermodynamically
unfavorable due to the dominance of entropy over enthalpy. \label{fig:dhT}
}
\end{figure}

The spinodal humidity range, which provides an upper bound on the humidity
hysteresis in this model, can be derived analytically:
 \begin{equation} \label{eq:dh1}
  \Delta h_{sp} = \frac{2}{c_0} \, f\left(\frac{T_c}{T}\right)
  \exp\left(-\frac{\Delta q_a}{k_B T}\right)
 \end{equation}
where the Arrhenius temperature dependence is augmented by a prefactor
 \begin{equation}
  f(u) = \frac{c_T \, \Delta h_{sp}}{2}= (2u-1)\sinh v - v \cosh v, \ \mbox{ where } \
  v = 2\sqrt{u(u-1)}, \ u = \frac{T_c}{T} = \frac{\omega}{2 k_B T}.
 \label{eq:dh2}
\end{equation}
An important prediction of this model is that {\it sorption hysteresis
decreases with increasing temperature and vanishes as a power law at the
critical point:}
 \begin{equation}
  \Delta{h}_{sp} \sim \frac{8}{3c_T} \left( \frac{T_c}{T} - 1 \right) ^{3/2} \
  \mbox{ as } T \to T_c, \label{eq:power}
 \end{equation}
The same $\frac 3 2$ critical exponent also arises in the temperature
dependence of the interfacial tension between the low-density and
high-density phases in the van der Waals 
theory of capillarity~\cite{vdw1893} and the related Cahn-Hilliard model of
phase separation~\cite{CahHil58}, which we consider in the next section.
(In structural mechanics, this scaling relation is analogous to Koiter's
2/3-power law for the difference between the critical load at symmetric
bifurcation of a perfect structure and the stability limit of imperfect
structures with vanishing imperfections~\cite{BazCed91}. There is also
Koiter's 1/2-power law for the bifurcation of a perfect system that is
asymmetric, and we would expect  analogous scaling, $\Delta{h}_{sp} \propto
(T_c - T)^2$ in  a different model with broken symmetry in the entropy
and/or enthalpy density around $\Theta = \frac{1}{2}$.) More generally, the
theory of critical phenomena provides many ways for nontrivial power law
scalings to arise, $\Delta{h}_{sp} \propto (T_c - T)^\nu$, and the exponent
$\nu$ is best determined by experiment for a given material.

As shown in Figure ~\ref{fig:dhT}, the critical power law (\ref{eq:power})
is a good approximation of the exact formula (\ref{eq:dh1})-(\ref{eq:dh2})
over a broad temperature range corresponding to typical hysteresis values,
$\Delta h < 10$.  In these formulae, the weak dependence of $c_T$ on
temperature can be neglected if orthogonal adsorption forces are much
stronger than lateral intermolecular forces, $\Delta q_a \gg 2 k_B
T_c=\omega$, which is always the case whenever there is significant
adsorption from the vapor. (Otherwise, bulk liquid condensation would occur
before surface adsorption.)  For water adsorption in cement   
paste or concrete, this assumption is consistent with an early estimate of
$\Delta q_a/k_B = \Delta Q_a/R = 2700$ K  ~\cite{bazant1972}, which is
likely to be much larger than $T_c$, as discussed below (see
also~\cite{BazKap96}, p. 210).

In such cases, the  dominant temperature dependence in (\ref{eq:dh1}) comes
from the prefactor (\ref{eq:dh2}), which vanishes at the critical point.
Physically, sorption hysteresis disappears above the critical temperature
because entropy, which promotes uniform surface coverage, then dominates
the enthalpy of lateral interactions, which promotes molecular
condensation. This is a very general effect, which will also arise in more
complicated models (e.g. for multilayer adsorption
~\cite{seri-levy1993,Nik96,CerMed98,CerMed98b}), as long as the lateral
interactions among adsorbed molecules are attractive.

\subsection*{ Molecular Condensation on Bounded Surfaces }

 {\bf Interphasial Tension: }
The foregoing theory describes sorption hysteresis for an infinite pore or
surface, since it considers only the homogeneous free energy per site. In a
finite system, phase separation is hindered by the interfacial (or
``interphasial") tension between immiscible stable phases, below the
critical temperature. The standard mathematical model of interphasial
tension is based on the concept  of a diffuse interface of continuously
varying density,  first introduced by Van der Waals in his original
``thermodynamic theory of capillarity"~\cite{vdw1893} and still used today
to describe surface wetting by thin liquid
films~\cite{degennes1985,cahn1977,widom1999,gouin2008}. The same model has
also been used to describe disjoining pressure in liquid-filled
nanopores~\cite{gouin2009}, albeit without making connections to adsorption
isotherms and sorption hysteresis in nanoporous solids. Modern interest in
this approach and many subsequent extensions sprang from the celebrated
paper of Cahn and Hilliard~\cite{CahHil58}, which used the regular solution
model to rederive and extend key results of Van der Waals~\cite{vdw1893}
and paved the way for phase-field models in materials
science~\cite{thornton2003,baluffi}.

Given the homogeneous free energy (\ref{eq:ghom}) per discretized pore
volume $V_s(\vec{r})$ (associated with fixed sites on the nearby pore
surface), the total free energy $G$ of an arbitrary (possible multiply
connected) pore geometry can be expressed as an integral over the pore
volume,
 \begin{equation}   \label{eq:CH}
  G[\Theta,S] = \int d\vec{r} \left[ g_{hom}(\Theta,\vec{r})
  + \frac{1}{2} \nabla\Theta\cdot \KK(\vec{r}) \nabla\Theta \right]
 \end{equation}
which is a functional of $\Theta(\vec{r})$, the dimensionless filling
fraction of the volume $V_s(\vec{r})$ (e.g. measured in monolayers, and
possibly larger than one for a site volume spanning a nanopore) and
$S(\vec{r})$, a function prescribing the surface geometry of the pore (e.g.
via a level-set or phase-field description). The coefficient $\KK$ in the
second term is the ``gradient penalty tensor", which approximates
corrections to the free energy density due to density variations, as well
as the interphasial tension (see below). In principle, both $g_{hom}$ and
$\KK$ depend on position within the pore geometry specified by $S$.  In Eq.
(\ref{eq:CH}), we neglect the mechanical energy stored in the solid
phase~\cite{thornton2003,meethong2007b,cogswell2011}, in order to emphasize
our prediction that hysteresis can occur in nanoporous solids whose
mechanical deformation, if any, is too small to affect the equilibrium
distribution of the adsorbate. It would be straightforward to incorporate
mechanical response of the solid matrix in a more sophisticated model, e.g.
following Ref.~\cite{cogswell2011}.

The diffusional chemical potential is given by the functional derivative
with respect to composition,
 \begin{equation}     \label{eq:mu}
  \mu = \frac{\delta G}{\delta \Theta}  = \mu_{hom} - \nabla\cdot \KK \nabla \Theta
 \end{equation}
where $\mu_{hom} = g_{hom}^\prime$ is the homogeneous chemical potential
(\ref{eq:muhom}). Physically, this corresponds to the free energy change
upon creating a continuum ``molecule" represented by a delta function at
$\vec{r}$. Outside the spinodal range, setting chemical potential $\mu=$
constant yields one uniform solution, corresponding to a stable homogeneous
phase. Within the spinodal range, there are three uniform solutions, two
stable and one unstable, and $\mu=$ constant yields a Beltrami differential
equation, whose nontrivial solution corresponds to a phase separated system
with a diffuse phase boundary. In the regime of strong phase separation
$\omega \gg k_BT$ or $T \ll T_c$, the width $\lambda$ of the phase boundary
scales as
 \begin{equation}
  \lambda =  \sqrt{\frac{K}{\omega}}
 \end{equation}
in each eigendirection of the $\KK$ tensor, and the corresponding
interphasial tension (energy/area) is
 \begin{equation}
  \gamma = \rho_s \sqrt{K \omega}
 \end{equation}
where $\rho_s$ is the density of sites per volume, or the inverse of the
single-site volume~\cite{CahHil58,burch2009}.

 {\bf Suppressed Condensation in Small Pores: }
The tendency for phase separation is reduced in small systems, as the phase
interface area to bulk volume ratio increases. More precisely, both the
spinodal range~\cite{cahn1961} and miscibility
gap~\cite{nauman1989,nauman1991,burch2009} shrink and ultimately disappear,
as the system size becomes comparable to the phase boundary thickness. For
phase separation in solid materials, if the two phases have different
equilibrium volumes, then elastic coherency strain energy further reduces
the miscibility and spinodal gaps and can eliminate phase
separation~\cite{cogswell2011}.

The suppression of phase transformations with decreasing system size is a
universal phenomenon, which is drawing attention in other fields and has
important technological applications. For example, it controls the
``ultimate fineness", or minimum feature size, of polymer-in-polymer
microdispersions~\cite{nauman1989,nauman1991}. Recently, it has also played
a major role in the development of  high-rate Li-ion batteries using
lithium iron phosphate (Li$_x$FePO$_4$), which has a strong tendency for
phase separation into Li-rich and Li-poor domains. Ironically, when it was
first explored as an insertion cathode material in microparticle form,
Li$_x$FePO$_4$ was predicted to be good for "low power applications" as a
result of slow phase separation dynamics and related mechanical
deformations~\cite{pahdi1997}, but today, in nanoparticle form, it is
capable of ultrafast battery discharge (in tens of seconds) while
maintaining long cycle life~\cite{kang2009}. In addition to size-dependent
diffusivity~\cite{malik2010}, the main reasons that ``nano is different"
may be the shrinking of the miscibility gap
\cite{meethong2007a,meethong2007b,burch2009,wagemaker2011,cogswell2011} and
the dynamical suppression of phase
separation~\cite{malik2011,bai2011,cogswell2011}. Analogous phenomena must
also occur for vapor sorption in nanopores, as we now explain.

Consider a homogeneous adsorbate $\Theta=\Theta_0$ in a straight pore or
flat surface whose longest lateral dimension is $L$. For example, in the
case of a perfect cylindrical pore, we set $L$ equal to the maximum of its
length and diameter. This state will be linearly stable to sinusoidal
perturbations of wavevector $\vec{k}$, given by $\Theta(\vec{r}) =
\Theta_0(1 + \epsilon e^{i \vec{k}\cdot\vec{r}})$, if the perturbation
increases the chemical potential, which implies
 \begin{equation}
  \frac{\partial \mu_{hom}}{\partial \Theta}\left(\Theta_0\right)
  + \vec{k}\cdot\KK\vec{k} > 0
 \end{equation}
The second term is strictly positive, so in an infinite system, where
arbitrarily long wavelength perturbations ($k\to 0$) with vanishing
gradient energy are possible, the spinodal range is defined by setting the
first term to zero, as above. In a finite system, however, there is a
minimum wavelength for perturbations set by the boundary conditions, e.g.
given by $k_{min}=\frac{\pi}{L}$ for constant concentration boundary
conditions, reflecting adsorption equilibrium at the farthest ends of the
pores. As a result, the spinodal range of unstable compositions, determined
by $\mu<0$, is generally reduced~\cite{cahn1961,nauman1989,burch2009}.

The corresponding spinodal humidity range $\Delta h_{sp}(L)$, defined as
the jump in the homogeneous isotherm humidity between the spinodal points
satisfying $\mu(\Theta,L)=0$, is given by the same formula as derived
above, Eqs. (\ref{eq:dh1})-(\ref{eq:dh2}), but only with a length-dependent
critical temperature,
 \begin{equation} \label{eq:TcL}
  T_c(L) = T_c^\infty \left( 1 - \frac{\pi^2 \lambda^2}{2 L^2} \right), \
  \mbox{ where } \ T_c^\infty = \frac{\omega}{2k_B}.
 \end{equation}
The critical point is depressed as the ratio of the system size to the
phase boundary thickness $\frac{\lambda}{L}$ decreases. For very small
systems, $L< L_c = \frac{\pi \lambda}{\sqrt{2}}$, the critical temperature
vanishes, and the homogeneous state is linearly stable, even at zero
temperature. In simple physical terms, molecular condensation is only
possible on surfaces whose length $L$ is large enough to accommodate an
equilibrium phase boundary of thickness $\lambda$.

\begin{figure*}
\begin{center}
\vspace{-.2in}
\includegraphics[width=5in]{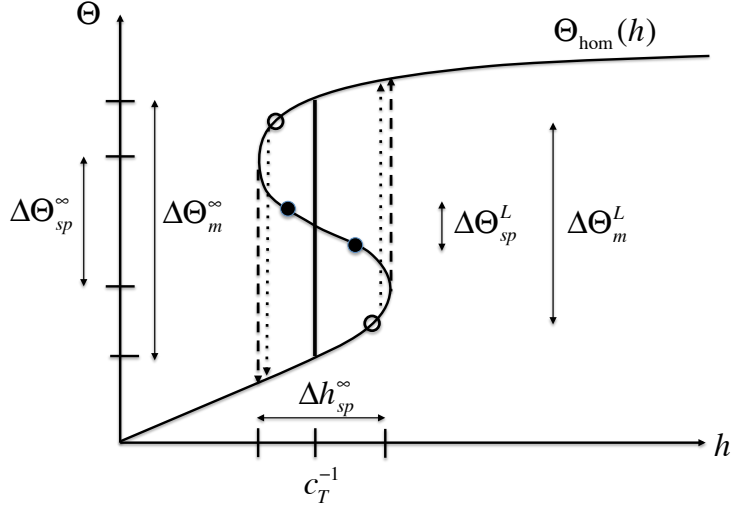}
\end{center}
 \vspace{-1in}
\caption{ Molecular condensation on a bounded surface of maximum dimension
$L$ (e.g. length or diameter of a cylindrical pore), where humidity $h$ is
controlled (rather than the mean filling fraction $\Theta$). If nucleation
is possible, then, as the miscibility gap $\Delta \Theta_m^L$ shrinks (open
circles) and the nucleation rate decreases with decreasing pore size, the
hysteresis (dotted lines) increases up to the maximum set by the spinodal
humidity range, $\Delta h_{sp}^\infty$. (b) In the absence of nucleation,
the adsorbate undergoes sudden transitions at the infinite-system spinodals
(dashed lines), prior to reaching the reduced spinodal gap
$\Delta\Theta_{sp}^L$ (black points). \label{fig:short} }
\end{figure*}

 {\bf Enhanced Hysteresis in Small Pores: }
One might expect hysteresis to be suppressed in short pores, due to the
reduced spinodal and miscibility gaps, but this is not the case since it is
the humidity, not the filling fraction, that is experimentally controlled.
The basic physics is sketched in Figure ~\ref{fig:short}. First consider
the possibility of nucleation, where the second phase is created by
fluctuations over the surface or at defects or pore edges. As the pore size
is decreased, there are fewer sites for nucleation, and the reduced
nucleation probability enhances hysteresis by preserving the homogeneous
state as the humidity is varied. Even if nucleation is very fast, the
reduced miscibility gap $\Delta h_{hom}^L$ increases the corresponding
humidity range of hysteresis, up to the limit set by $\Delta
h_{sp}^\infty$.

In the absence of nucleation, phase separation must occur by spinodal
decomposition, which is also suppressed in short pores. If the humidity
passes out of the spinodal range, then the adsorbate can pass into the
infinite-system spinodal  gap $\Delta h_{hom}^\infty$ while remaining
uniform. After the overshoot, it experiences a strong thermodynamic driving
force to vary the concentration until the new homogeneous equilibrium state
is reached, and phase separation may not have enough time to occur during
this sudden transition. (This 
non-equilibrium transition state has recently been called a ``quasi-solid
solution"~\cite{bai2011}.)  As a result, the infinite-system spinodal
humidity range, $\Delta h_{sp}^\infty(T)$, given by Eqs.
(\ref{eq:dh1})-(\ref{eq:power}), provides a robust estimate of sorption
hysteresis in finite-length pores of any size if nucleation is too slow to
occur over experimental time scales.

\subsection*{ Molecular Condensation in Single Nanopores }

 {\bf  Theory of Hindered Multilayer Adsorption: }
A general continuum model of adsorption in nanoporous media of arbitrary
geometry could be based on the van 
der Waals (or Cahn-Hilliard) model, Eq. (\ref{eq:CH}), where
$\Theta(\vec{x})$ is the local mean density of the adsorbate at a point in
the pore. The main difference with monolayer adsorption is that the
adsorption free energy $\Delta g_a(\vec{x})$ would depend on $\vec{x}$ and
reflects the decay of surface forces with distance from the pore walls,
including screening effects due to the other molecules. The local
interaction energy, $\omega(\vec{x})$, would also depend on position,
approaching a bulk value as the influence of surface forces decays with
distance from the nearest wall. The gradient penalty $K(\vec{x})$ would
represent the local free energy difference associated with broken or
frustrated molecular bonds (the nanoscale analog of gas-liquid surface
tension), and this, too, would generally depend on position.

The dynamics of the concentration profile in such a model would be
described by the Cahn-Hilliard equation \cite{baluffi,naumann2001},
 \begin{equation}
  \frac{\partial \Theta}{\partial t} = \nabla\cdot \left[ M \Theta \nabla \mu \right]
 \end{equation}
where the chemical potential $\mu(\Theta,\vec{x})$ is given by
(\ref{eq:mu}) and the diffusional particle mobility $M(\Theta,\vec{x})$
generally depends on the concentration and position. For example, in the
foregoing regular solution model 
for the first monolayer, the mobility should be proportional to the free
volume, $M = M_0 (1- \Theta)$ (this effect was omitted in early models and
yields a ``modified Cahn-Hilliard equation"~\cite{naumann2001}).   More
generally, to account for finite pores, one should use the
``Cahn-Hilliard-Reaction model", which includes thermodynamically
consistent boundary conditions for molecules to enter and leave the pore
space \cite{singh2008,burch2009}. It is beyond the scope of this paper to
analyze or simulate the intrapore adsorbate distribution in detail, but
this would be interesting to pursue in future work. Such a nanoscale,
statistical continuum model may be able to capture key features of
molecular dynamics
simulations~\cite{gelb1999,Pel-Lev02,Coa-Pel07,Coa-Pel08,Coa-Pel08b,Coa-Pel09}
at greatly reduced computational cost, thereby allowing extensions to
experimental time and length scales.

 {\bf Hierarchical Wetting Model: }
We proceed instead by making some simple approximations to enable us to
predict general features of molecular condensation in nanoporous media. It
is often reasonable to assume that in the first monolayer there are strong
surface forces, which decay quickly with distance from the surface. In case
of concrete, for example, water adsorption in C-S-H nanopores containing
dissolved salts involves strong electrostatic forces, which are screened at
the molecular scale due to diffuse charge, solvated ion crowding and
electrostatic correlations~\cite{pellenq1997,RTIL2011}. For simplicity, let
us assume that the regular solution model above describes the mean
homogeneous coverage $\Omega_{hom}(h,T;c_T,\omega)$ of the first monolayer
on the surface at low vapor pressure. Here, $c_T$ and $\omega$ describe the
local adsorption and interaction energies, which could depend on surface
heterogeneities or curvature in very small pores. In thick pores at low
vapor pressure, the dynamics of the concentration profile in the first
monolayer could be described by the Allen-Cahn equation~\cite{baluffi} for
an open system (or its generalization for nonlinear adsorption
kinetics~\cite{bai2011}), since gas molecules are freely exchanged with the
adsorbate at all points.

Here we consider what would happen in the general case of nanopores, which
are thick enough to be covered by non-overlapping monolayers at low vapor
pressure, but thin enough to be spanned by a condensed liquid phase at
moderate vapor pressures (below the saturation pressure). Over experimental
time scales (e.g. minutes to months), the Cahn-Hilliard equation predicts
that small clusters of molecules or voids resulting from thermal
fluctuations of the homogeneous state will coarsen within the nanopores, so
as to minimize the gradient energy (which is the molecular analog of
surface tension). Assuming strong wetting in the first monolayer, this
coarsening proceeds until the size of the separated nanophases is set by
the diameter of the bulk region just outside the first monolayer.

\begin{figure*}
\begin{center}
\vspace{-.1in}
\includegraphics[width=6in]{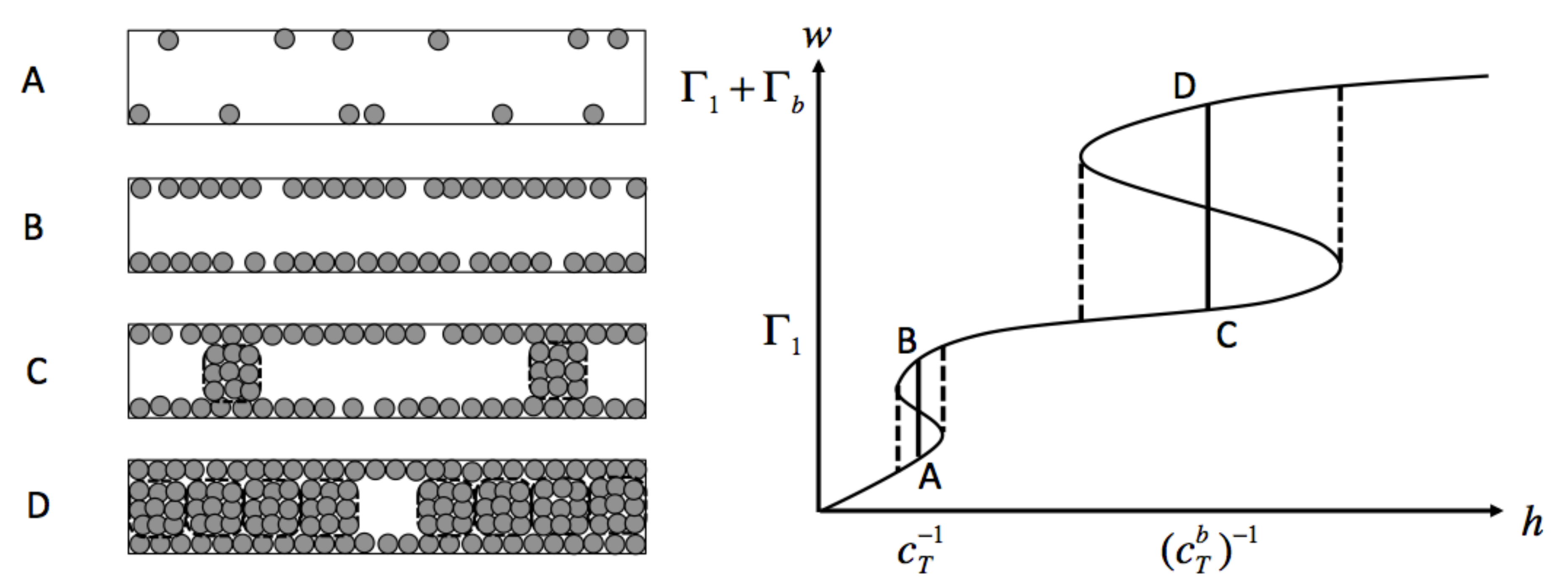}
\end{center}
\caption{ Hierarchical wetting model for hindered multilayer adsorption in
nanopores.  The homogeneous isotherm, as in Eq. (\ref{eq:sw}), exhibits two
regions of hysteresis: (i) a small loop  at low vapor pressure for
low-density (A) and high-density (B) phases in the first monolayer and (ii)
a larger loop  at higher vapor pressure for the low-density (C) and
high-density (D) phases of coarsened pore-spanning droplets and bubbles in
the bulk fluid, outside the monolayer. \label{fig:bubbles} }
\end{figure*}

In that case, as a crude first approximation, we describe the bulk region
by another regular solution model, whose characteristic lattice size is
that of the pore bulk. The filling fraction of the bulk region is thus
given by $\Theta_{\hom}(h,T;c^b_T,\omega^b)$, where $c_T^b = \exp(\Delta
g_a^b/k_B T)$ and $\omega^b$ are effective values for molecular droplets in
the pore. The total weight of a homogeneous adsorbate in the pore is then
given by
 \begin{equation}    \label{eq:sw}
  w_{hom}(h,T) = \Gamma_1 \Theta_{hom}(h,T;c_T,\omega)
  + \Gamma^b \Theta_{\hom}(h,T;c^b_T,\omega^b).
 \end{equation}
where $\Gamma_1$ is the total weight 
of a full surface monolayer and $\Gamma^b$ is the weight of the filled bulk
region.  This ``Hierarchical Wetting" model is surely oversimplified, but
it captures typical results of molecular dynamics
simulations~\cite{Coa-Pel07,Coa-Pel08,Coa-Pel08b,Coa-Pel09} and allows
considerable insight into experimental data, as we now explain.

 {\bf Surface versus bulk phase separation in nanopores: }
This very simple model leads to two types of hysteresis, one due to the
condensation of adatom clusters in the first monolayer at low vapor
pressure and the second due to the condensation of pore-spanning clusters
of molecules or voids (``nanodroplets" and ``nanobubbles", respectively) in
the bulk pore at higher vapor pressure.  Each hysteresis has its own
magnitude and critical temperature. If the pore radius is $R$ and monolayer
thickness, $a$, then the effective interaction parameter scales with the
geometrical ratio:
 \begin{equation}
  \frac{\omega^b}{\omega} \approx \frac{T_c^b}{T_c}
  \approx \frac{\Delta h^b}{\Delta h} \approx \frac{R}{a},
 \end{equation}
which is typically larger than one, except in molecular scale pores (which
might not be macroscopically accessible). An interesting implication is
that critical temperature $T_c^b$ for droplet phase separation in the bulk
nanopore is larger from that of the first monolayer $T_c$ by the same
factor $\alpha$.  Well below the critical temperatures, the magnitude of
the bulk hysteresis $\Delta h^b$ (dominated by enthalpy) is also larger
than that of the surface layer $\Delta h$, by the same factor. Bulk
hysteresis is also shifted to larger humidities in the multilayer
adsorption regime, assuming that the adsorption energy in the bulk is much
less than it is at the surface, $\Delta g_a^b \ll \Delta g_a$.

These simple concepts are illustrated in Fig. \ref{fig:bubbles}. As the
nanopore is emptied and filled, there are two instabilities corresponding
to sudden phase separation in the first monolayer, at low vapor pressure,
and in the bulk, at higher vapor pressure. The resulting desorption and
sorption isotherms for this crude model already resemble the experimental
data for concrete or cement paste at room temperature in
Fig.~\ref{fig:expt}. The theory also qualitatively predicts a nontrivial
effect of temperature, which suppresses hysteresis in the monolayers at low
vapor pressures more than in the bulk pores at higher vapor pressures.

 {\bf Snap-Through Instabilities: }
The reader may be wondering how our general statistical thermodynamic
theory in Part II relates to the snap-through instabilities predicted for
nonuniform pore openings in Part I. One major difference is that the theory
of Part I does not account for the entropy of molecular rearrangements in
the adsorbate and thus is mainly relevant for low temperature, where
enthalpy dominates and sorption or desorption proceeds sequentially through
the junction, like a crack tip. In the present model, phase separation can
occur anywhere in the system that achieves a locally metastable state, and
thus the sorption or desorption process can effectively tunnel through a
junction. Nevertheless, a junction between two nanopores of different
geometry or surface chemistry can act as a barrier for snap-through
instabilities, due to the interfacial  
tension associated with lateral interactions across the junction.

To model the pinning effect of molecular interactions at junctions, we use
again the Cahn-Hilliard (or Van der Waals) model (\ref{eq:CH}), but it
suffices to average over the cross section and focus on density gradients
through the junction or pore opening. Relative to the free energy of an
infinite system, the gradient energy per area (i.e. the interphasial
tension at the junction) can be approximated as
\begin{equation}
\Delta G_i \approx \lambda K \left( \frac{\Delta \tilde{\Theta}}{\lambda}\right)^2
\approx \omega^b \, \Delta \tilde{\Theta}^2  \label{eq:dG}
\end{equation}
where $\lambda$ is the local interface width and $\tilde{\Theta} =
\Theta/\Theta_{max}$ is the jump in dimensionless adsorbate density, or
total filling fraction ($0<\tilde{\Theta}< 1$). This additional energy
barrier must be overcome for phase separation to occur across the junction.
Just as in the foregoing case of short pores, 
the interphasial tension can lead to hindrance or even suppression of phase
separation.

\begin{figure}
\begin{center}
\includegraphics[width=7.5in]{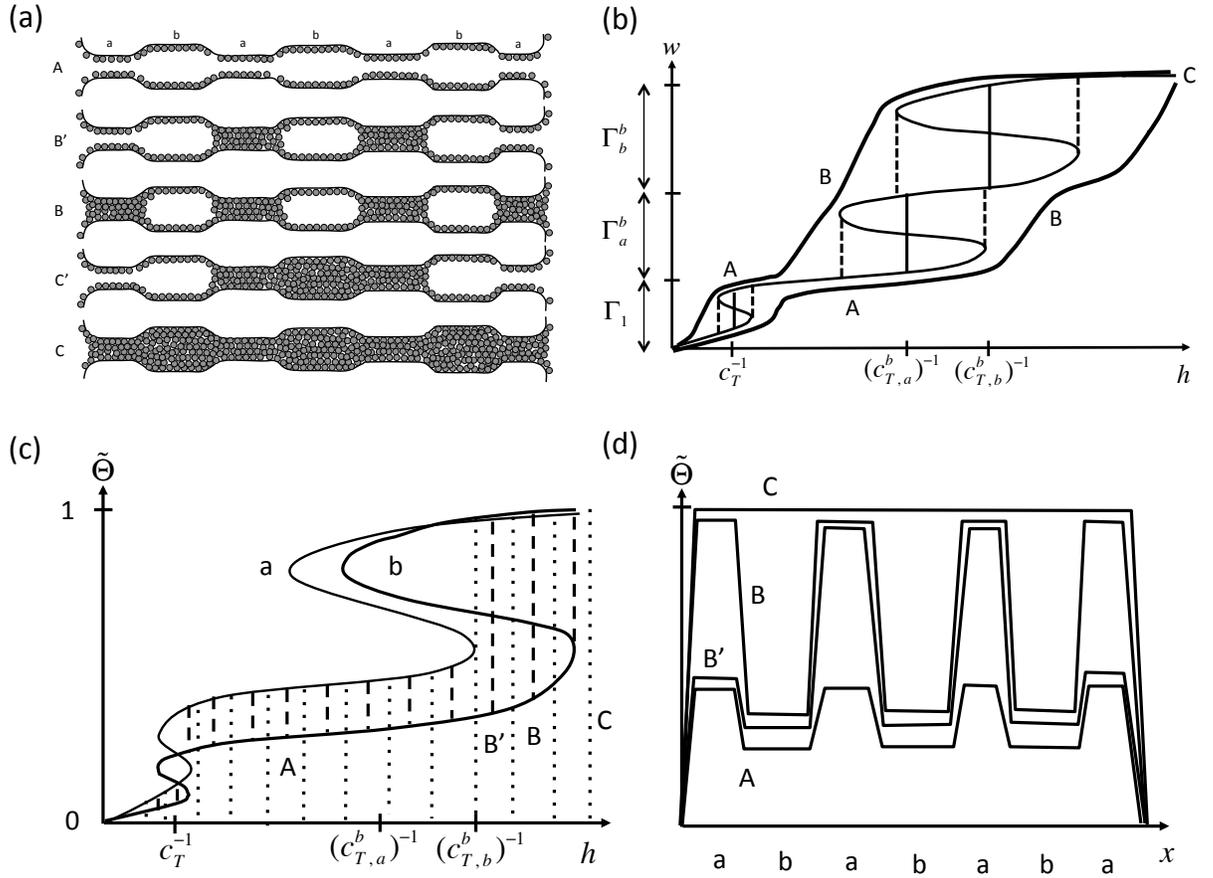}
\end{center}
\vspace{-.5in} \caption{ \label{fig:junction} Molecular condensation in a
heterogeneous multilayer nanopore with sections $a$ and $b$ of two
different thicknesses. (a) Typical molecular configurations A, B, B', C, and C' at
different stages of sorption. (b) Net sorption and
desorption isotherms for the nanopore including interfacial energies 
with the states A, B, and C indicated.
(c) Isotherms of dimensionless filling
fraction $\tilde\Theta$ (mean density) versus relative humidity $h$ for the $a$ and  $b$ sections of the pore, showing the jumps at $a$/$b$ junctions 
(dashed lines) and the empty regions at the pore ends (dotted lines).
(d) Spatial density profiles, $\tilde\Theta(x)$, for the states in part (a), where the interfacial energy due to each jump, which enhances  the hysteresis in (b), scales as $\omega^b \Delta\tilde\Theta^2$.  }
\end{figure}

When phase separation does occur across the junction, the chemical
potential jumps according to Eq. (\ref{eq:mu}) and causes an increase in
humidity hysteresis at a given mean weight, due to the need to overcome the
interfacial energy at the junction.   As shown in Figure
~\ref{fig:junction} for the heterogeneous pore geometry in (a), this effect
can be estimated graphically using Eq. (\ref{eq:dG}) by plotting in (c) the
two homogeneous isotherms as the dimensionless filling fraction
$\tilde{\Theta}$ versus humidity $h$ and measuring the density difference
$\Delta \tilde{\Theta}$ between the two curves as the humidity is varied.
The filling fraction jumps are also evident in the density profiles across
heterogeneous pore in (d).  The interfacial  contribution to hysteresis at
each junction, $\Delta G_i$, which leads to the enlarged hysteresis in (d), 
can be of comparable magnitude to the intrinsic hysteresis for the interior
of the pore, $\omega^b$, since each results from the energy of lateral
forces in a cross section of the pore. As the interfacial energy builds up during sorption, the humidity is gradually amplified by $\exp(\Delta G_i/k_BT)$, while during desorption the humidity is multiplied by $\exp(-\Delta G_i/k_BT)$. The net effect is to widen the hysteresis envelope associated with each phase separation process, e.g. in the first monolayer and the two types of bulk pores.

The sorption/desorption sequences predicted by these arguments using the
simple Hierarchical Wetting Model are similar to those observed in
molecular dynamics simulations of wetting fluids in heterogeneous nanopores
by Coasne, Pellenq and
collaborators~\cite{Coa-Pel07,Coa-Pel08,Coa-Pel08b,Coa-Pel09}. The example
shown in Figure ~\ref{fig:junction}(a) consists of a series of
multilayer-thick pores of two different radii $a$ and $b$, which terminates
at the free surfaces of much larger pores. The bulk regions of the thicker
$b$ pores have smaller adsorption energy $\Delta g^b$ (larger $c_T^{-1}$)
and larger interaction energy $\omega^b$ for spanning nanodroplets and
nanobubbles than those of the thinner $a$ pores. This leads to the
different rescaled isotherms $\tilde \Theta(h)$ for $a$ and $b$ sections
plotted in (c), which allow the sorption/desorption sequences to be
predicted.  During sorption starting from very low humidity, a monolayer
first covers the entire pore surface in state A, and then bulk molecular
condensation proceeds to the narrower sections in state B, followed by the
thicker sections in state C. The corresponding spatial profiles of the
filling fraction are shown in (d), from which the interfacial energies at
the junctions can be estimated using Eq. (\ref{eq:dG}). Due to the larger
density jump at the pore ends, there is a larger interfacial energy there,
compared to the internal $a$/$b$ junctions, and this can cause condensation
to occur in outermost pores after the others, leading to intermediate
states B' and C' shown in (a).

 {\bf Solid Matrix Deformation: }
Although we emphasize the statistical thermodynamic origin of sorption
hysteresis, mechanical deformation of the solid matrix during phase
separation could play an important role, even in the absence of pore
collapse. Molecular condensation events lead to sudden changes in
disjoining (or joining) pressure, which can influence neighboring pores,
perhaps even triggering chains of subsequent phase separation events, as
stresses are quickly transmitted through in the solid (at the local speed
of sound). Analogous effects of ``coherency strain" due to adsorption in
elastic crystal lattices have recently been considered in Li-ion batteries
and shown to contribute to suppression of phase separation and nonlinear
pattern formation, consistent with experimental
observations~\cite{cogswell2011}. Ultimately, a complete model of sorption
dynamics in nanoporous solids should take into account viscoelastic
relaxation, or even plastic deformation and damage to the solid matrix,
coupled to molecular condensation and transport.

\subsection*{ Molecular Condensation in Nanoporous Solids }

\begin{figure*}
\begin{center}
(a)\includegraphics[width=3in]{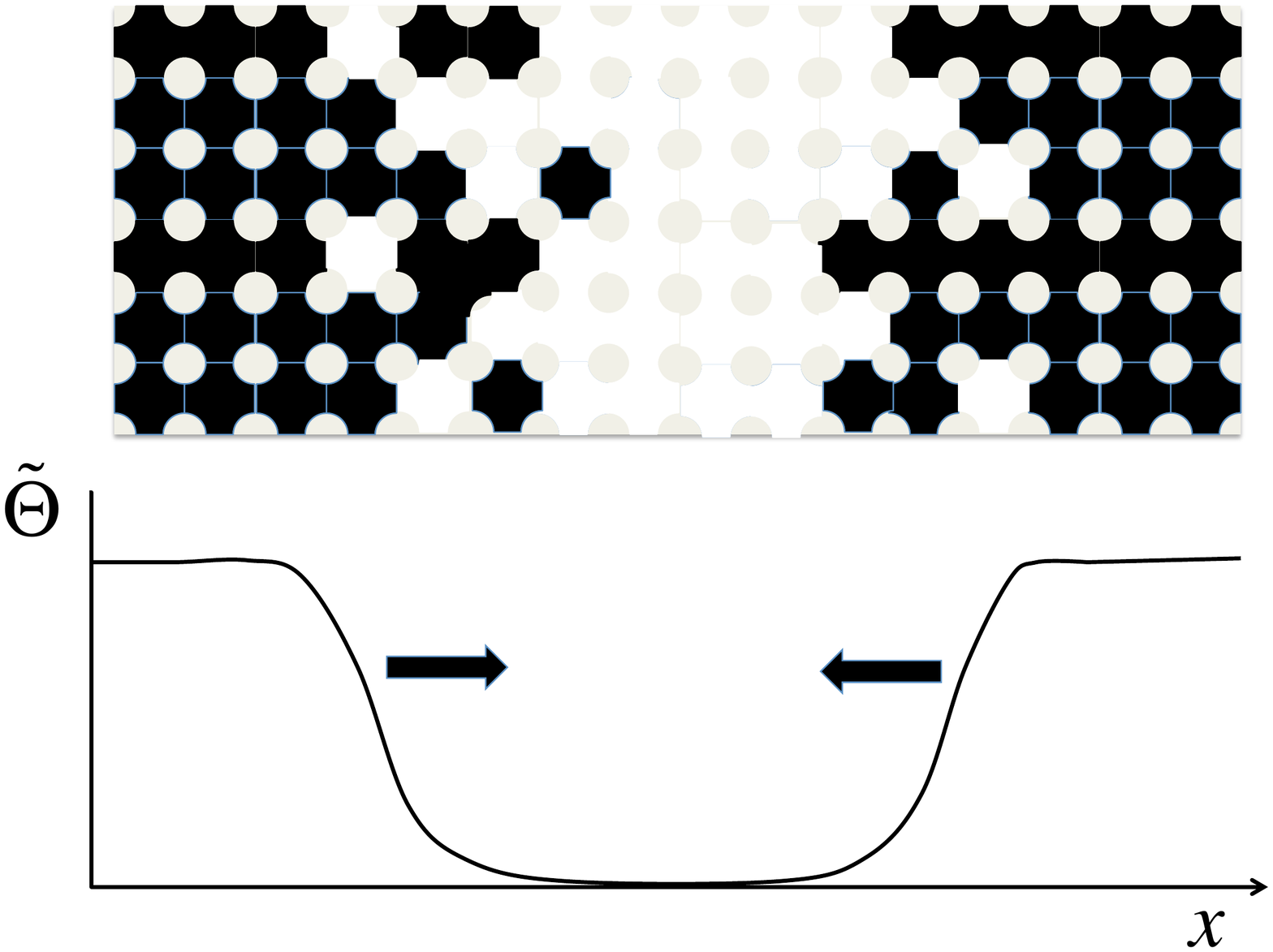}
(b)\hspace{-0.3in} \includegraphics[width=3.2in]{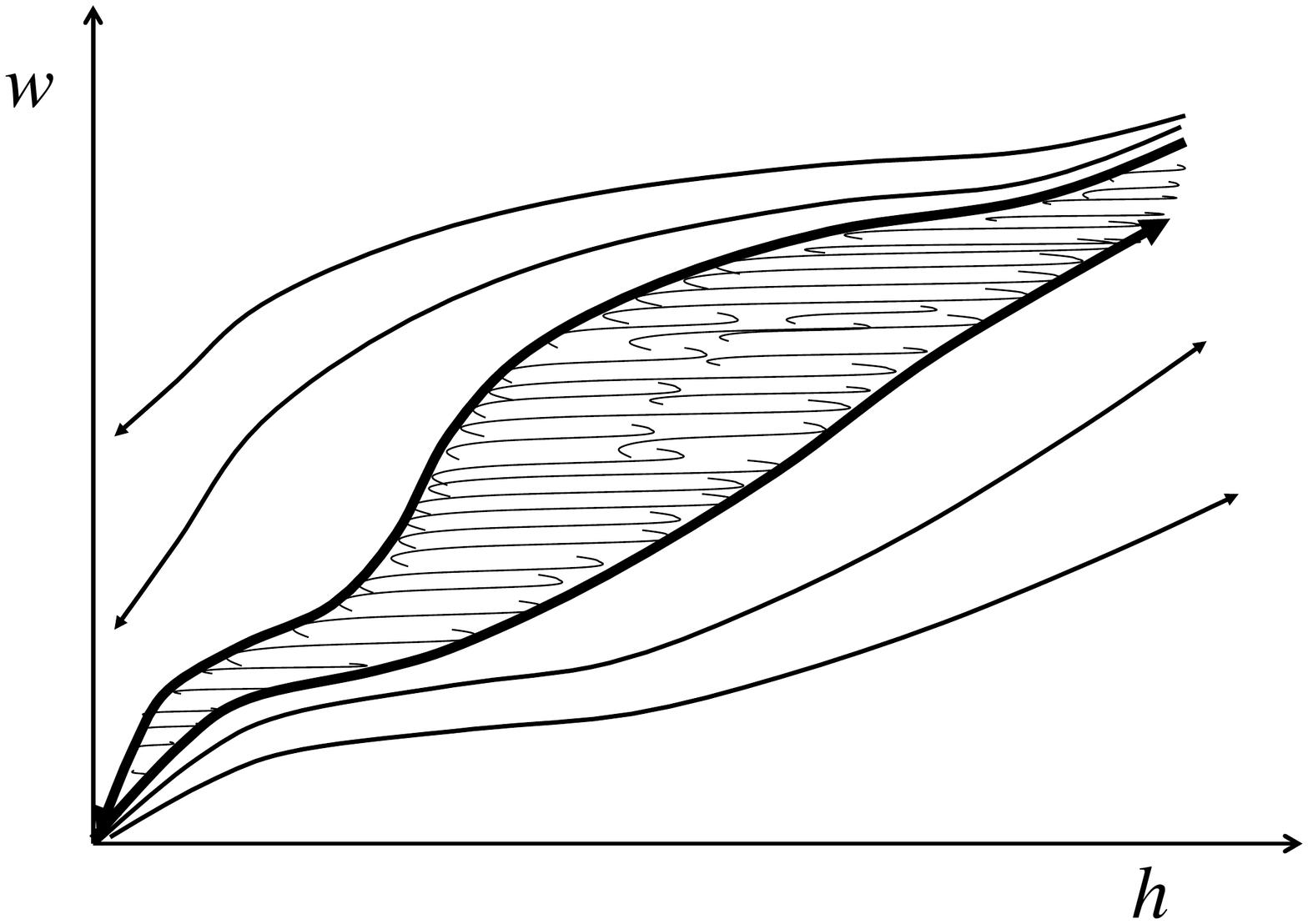}
\end{center}
\caption{ Molecular condensation in macroscopic nanoporous solids. (a) Discrete nanopore transformations due to local mosaic instabilities propagate across the material in a narrow front from the edges where the vapor pressure is controlled. The front thickness scales with sweep rate $dw/dt$, and at high rates spans the system. (b) Sorption hysteresis due to quasi-equilibrium thermodynamics (thick solid lines) corresponding to countless, tiny condensation events (thin interior curves) for individual nanopores. With increasing sweep rate, there is additional non-thermodynamic hysteresis (thin exterior curves) due to internal resistances to transport and/or adsorption reaction kinetics.
\label{fig:mosaic}
}
\end{figure*}

 {\bf Mosaic Instability: }
It is important to emphasize that quasi-equilibrium phase separation across
a nonuniform network of pores need not be sequential, as suggested in Part
I based on mechanical analogies. As long as there is sufficient time for
the chemical potential to equilibrate across the pore network between tiny,
applied steps in humidity, then there is no prescribed order for the
intermittent events of filling or emptying in different pores. At the
macroscopic scale, the resulting ``mosaic instability" of discrete
molecular condensation events in different, tiny nanopores manifests itself
by accumulating many incremental sorption/desorption isotherms into a
smoother overall curve, as shown in Figure~\ref{fig:mosaic}. This effect
has recently been invoked to explain the noisy thermodynamic hysteresis of
voltage versus state of charge in phase-separating nanocomposite Li-ion
battery cathodes ~\cite{dreyer2010,dreyer2011}.

This approximation inevitably breaks down, however, in sufficiently large
porous bodies, 
where the transport of mass or heat is too slow to fully equilibrate the
system between external humidity steps. In such cases, the progression of
phase separation is biased by transport limitations and indeed behaves like
a sequence of snap-through instabilities propagating through the porous
medium, starting from the edges where the humidity is controlled. Analogous
phenomena have also recently been predicted ~\cite{ferguson} and
observed~\cite{harris2010} in Li-ion batteries, where a narrow front of
mosaic instability can propagate across the electrode from the separator,
limited by diffusion of lithium ions in the electrolyte. Similar dynamical
phase separation phenomena must occur during vapor sorption/desorption in
porous media, but to our knowledge no simple theory is yet available for
the resulting macroscopic dynamics. Percolation concepts, which have been
applied to sorption hysteresis due to classical capillarity in pore
networks~\cite{liu1993}, may be useful, but molecular condensation, gas
transport and adsorption reaction kinetics should also be considered.
Analyzing the macroscopic dynamics of sorption and desorption in phase
separating nanoporous solids would be a very interesting avenue for future
research.

 {\bf Rate Dependence. }
Non-equilibrium phenomena always play a role in hysteresis, no matter how slowly experiment is performed. Whenever transport
or adsorption kinetics are at least partly rate limiting, there will be an
additional non-thermodynamic contribution to hysteresis, sketched in Fig.
~\ref{fig:mosaic}(b), which is related to the work done (or frictional
energy dissipated) to drive the system, as noted in Part I. The faster the
humidity sweep rate, the larger the hysteresis, in excess of the
thermodynamic contribution described above. This effect is analogous to the
``overpotential" or ``internal resistances" observed in battery cycling
under different conditions. For example, galvanostatic discharge is
analogous to constant mass flux $dw/dt$, and more realistic situations for
gas sorption, such as small humidity steps $\Delta h$ or constant rate of
humidity variation $dh/dt$, are analogous to potentiostatic intermittent
titration and cyclic voltammetry, respectively. In principle, detailed mathematical
modeling of transient vapor sorption/desorption taking into account
transport, adsorption kinetics and phase separation could enable
quantitative information about the material to be extracted from the
rate-dependence of the observed hysteresis.

\begin{figure*}
\begin{center}
\vspace{-.7in}\includegraphics[width=5in]{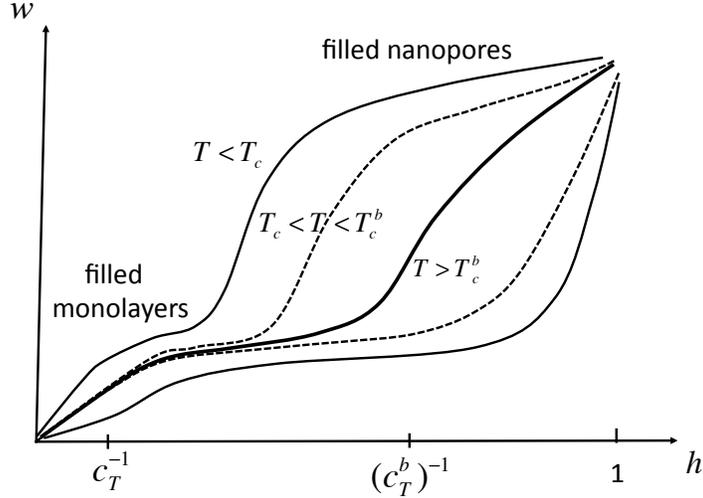}
\end{center}
\vspace{-0.5in}
\caption{ Temperature dependence of sorption hysteresis in nanoporous solids with strong wetting. At low temperature, pronounced hysteresis exists at all vapor pressures (thin solid curves). Above the critical temperature $T_c$ for phase separation in a monolayer, hysteresis disappears at low vapor pressure, but remains at high vapor pressure (dashed curves) until the critical temperature for bulk nanopore phase separation, $T_c^b$, is exceeded (thick solid curve).   \label{fig:cycles_temp}
}
\end{figure*}

{\bf Temperature Dependence. }  As noted above, the very simple Hierarchical Wetting Model already makes an interesting prediction about the temperature dependence of hysteresis in nanoporous solids with strong wetting. As sketched in Figure~\ref{fig:cycles_temp}, hysteresis is pronounced at low temperature and disappears at high temperature, but there is an intermediate range of temperatures ($T_c< T < T_c^b$) where hysteresis vanishes at low vapor pressure (monolayer filling) but persists at high vapor pressure (bulk nanopore filling). In complicated nanoporous geometries, the general effect should remain: As the temperature is increased, hysteresis vanishes first at low vapor pressure in the monolayer regime and then at high vapor pressure in the bulk nanopore regime. The reason is that the effective interaction energy $\omega^b$ of pore-spanning nanodroplets and nanobubbles is larger than that of individual adsorbed molecules, due to the larger number of intermolecular bonds. Similar arguments apply to the interfacial energy $\Delta G_i$ at a nanopore junction, which is larger at high filling than for monolayer coverage. As a result, the critical hysteresis temperatures for different bulk nanopores are larger than for monolayers.

\subsection*{ Re-interpretation of Experimental Data for Concrete: }

As noted above, water sorption hysteresis at low humidity in concrete has
long been attributed to pore collapse, in spite of the lack of any testable
theory.  It is noteworthy, therefore, that a number of puzzling
experimental observations can be explained for the first time by molecular
condensation without invoking any changes in the pore structure. In some
cases, our predictions seem to be quantitatively consistent with the data,
although more systematic experiments and detailed modeling should done,
e.g. at different temperatures and humidity sweep rates, to further test
the theory.

\begin{itemize}

 \item {\bf Inert gas versus water vapor.}
No sorption hysteresis is observed for inert gases, such as nitrogen, in
the same concrete samples that exhibit large hysteresis for water
vapor~\cite{baroghel2007}, and  our model is able to attribute this effect
to differences in lateral forces. Sorption experiments are usually carried
out at room temperature, which is larger than the critical temperature
$T_c$ in our model (\ref{eq:Tc}), if the mean pair interaction energy
$\omega$ is smaller than $2 k_B T = 52$
meV. This is a reasonable upper bound for the weak lateral van     
der Waals forces in adsorbed inert gases, and 
so negligible hysteresis can be expected. In contrast, adsorbed water
molecules have much stronger attractive forces, leading to room-temperature
hysteresis, as explain next.  

 \item {\bf Hysteresis at low vapor pressure.}
Our theory, although oversimplified, predicts hysteresis of a reasonable
scale for monolayer water sorption at room temperature.  Naively, we might
estimate $\omega$ by the hydrogen bond enthalpy in bulk liquid water of
23.3 kJ/mol~\cite{suresh2000}, which would imply $T_c = 1380$ K,  but this
grossly overestimates the lateral pair interaction energy of adsorbed water
molecules. A recent molecular dynamics study of water monolayers on
hydrophilic silica surfaces has shown that 90\% of the water molecules are
``non-wetting", having much stronger bonds with the surface than with other
water molecules~\cite{romero2011}. If we work backwards from our result in
Figure ~\ref{fig:dhT}, then 1\% hysteresis (in relative humidity) at room
temperature implies $T_c = 410$K  or $\omega=$6.8 kJ/mol, while 10\%
hysteresis implies $T_c = 660$K or $\omega=$11 kJ/mol. These are reasonable
lateral interaction energies for adsorbed water molecules, which could
quantitatively explain the observed hysteresis without any pore collapse.

 \item {\bf Hysteresis at moderate vapor pressure.}  In the regime of
multilayer adsorption, we predict that the scale of hysteresis $\Delta h^b$
is larger than in the monolayer regime $\Delta h$, very roughly scaling as
the ratio of bulk pore radius to the monolayer thickness. This is
consistent with the larger hysteresis that is always observed in the
multilayer region~\cite{baroghel2007}. It may even be possible to make
quantitative connections with pore geometry, since the C-S-H pores and
wetting layers in cement paste are indeed at the scale of 3--10 molecular
diameters, as suggested by the hysteresis ratio.

 \item {\bf Temperature dependence.}
We predict that sorption hysteresis should decrease with increasing
temperature, although we are not aware of any prior theoretical predictions
or systematic experimental studies of this effect. In a recent
study~\cite{baroghel2007}, the hysteresis of water sorption in concrete at
low vapor pressure (first monolayer) was negligible at 44$^\circ$C but
quite significant at $23^\circ$C, albeit in different concrete specimens,
as shown in Fig. ~\ref{fig:expt}(a) and (b) respectively. The first
adsorption isotherm was also steeper at 44$^\circ$C than at $23^\circ$C
(Figs 3a and 9a of \cite{baroghel2007}), suggesting that the free energy
barriers responsible for hysteresis were lowered in the former case.
Moreover, the curve of nearly reversible sorption/desorption at the higher
temperature (Fig. ~\ref{fig:expt}(b)) passes roughly through the center of
the hysteresis ``window" between the first sorption and desorption curves
(Fig. ~\ref{fig:expt}(a)) over a wide range of low humidities ($h<60\%$).
The data in Fig. ~\ref{fig:expt}(b) also makes the tantalizing suggestion
of lingering hysteresis in the multilayer regime at an elevated temperature
where hysteresis is already suppressed in the monolayer regime. This is
consistent with our prediction that the critical temperature for
condensation of larger droplets in nanopores is much larger than for individual adatoms
in the first monolayer, as shown in Figure.~\ref{fig:cycles_temp}.

 \item {\bf Cycling History Dependence: }
Although we do not claim a quantitative understanding, it makes perfect
sense in light of our theory that the first few sorption/desorption cycles
often exhibit strong history dependence, where hysteresis grows with time,
until a more reproducible path is achieved. This would naturally result
from trapped condensed phases (droplets or bubbles) in the adsorbate, which
may require nucleation to a larger perturbation to be released, e.g. from
defects, cracks, or chemical heterogeneities. As explained above in the
context of nanopore junctions, any heterogeneity can act as a pinning site
for the adsorbate, effectively removing mass and increasing internal
resistance during the initial sorption and desorption cycles. Of course,
the same behavior 
is observed in rechargeable batteries, where the first few cycles are often
very different from next hundred, e.g. due to the lithium trapping in
interfacial films or defect sites, which lead to an irreversible initial
capacity loss.

\end{itemize}

\subsection*{Conclusion}

We have developed a simple thermodynamic theory of sorption/desorption
hysteresis in nano porous solids, based on the concept of hindered
molecular condensation. The model makes a number of novel and testable
predictions that seem consistent with previously unexplained data for
cement paste and 
concrete, without postulating any pore collapse. Further experiments to
systematically study the effects of temperature, humidity sweep rate,
cycling behavior, etc., are proposed.  The theory is very general and could
be refined at the nanoscale and connected with macroscopic transport and
mechanical deformation. The possibility of making  
quantitative predictions directly from the microstructure may lead to new,
more accurate methods of determining the internal surface area and nanopore
width distribution directly from the observed hysteresis of sorption and
desorption isotherms. For concrete in particular, fruitful new directions
for experiments suggested by our theory would involve systematically
varying the temperature and humidity sweep rate during sorption/desorption
of water vapor.

 {\small {\sf
\vv \no {\bf Acknowledgments:} This research was funded by NSF Eager
Collaborative Grants 1153509 to MIT and 1153494  to Northwestern University. 
The MIT Concrete Sustainability Hub also provided some support for this part (MZB).
Preliminary work was also funded by NSF under Grant DMS-0948071 to MIT (MZB) and Grant CMS-0556323 to Northwestern University (ZPB) and by the U.S. DoT Grant 27323  provided through the Infrastructure Technology Institute of Northwestern University (ZPB).  The
authors thank Rolland J.-M. Pellenq for valuable comments and references.
 }}



\begin{thebibliography}{99}  \setlength{\itemsep}{-2mm}
     {\small  

\bibitem{part1} Ba\v zant, Z. P. and Bazant, M. Z., ``Theory of sorption
hysteresis in nanoporous solids: I. Snap-through instabilities", submitted. 
Online preprint 	arXiv:1108.4949v1 [physics.flu-dyn].

\bibitem{AdoSet96} Adolphs, J., and Setzer, M.J. (1996). ``A model to
describe adsorption isotherms." {\em J. of Colloid and Interface Science}
180, 70--76.

\bibitem{AdoSet-02} Adolphs, J., Setzer, M.J. and Heine, P. (2002).
``Changes in pore structure and mercury contact angle of hardened cement
paste depending on relative humidity." {\em  Materials and Structures} 35,
477--486.

\bibitem{Bal-Gub93} Balbuena, P.B., Berry, D., and Gubbins, K.E. (1993).
``Solvation pressures for simple fluids in micropores." {\em J. Phys.
Chem.} 97, 937--943.

\bibitem{Baz70} Ba\v zant, Z.P. (1970). ``Constitutive equation for
concrete creep and shrinkage based on thermodynamics of multi-phase
systems." {\em Materials and Structures}, 3, 3--36 (reprinted in {\em Fifty
Years of Evolution of Science and Technology of Building Materials and
Structures}, Ed. by F.H. Wittmann, RILEM, Aedificatio Publishers, Freiburg,
Germany 1997, 377--410).

\bibitem{Baz70b} Ba\v zant, Z.P. (1970). ``Delayed thermal dilatations
of cement paste and concrete due to mass transport." {\em Nuclear
Engineering \& Design}, 24, 308--318.

\bibitem{Baz72} Ba\v zant, Z.P. (1972). ``Thermodynamics of interacting
continua with surfaces and creep analysis of concrete structures." {\em
Nuclear Engineering and Design}, 20, 477--505.

\bibitem{Baz72a} Ba\v zant, Z.P. (1972). ``Thermodynamics of hindered
adsorption with application to cement paste and concrete." {\em Cement and
Concrete Research}, 2, 1--16.

\bibitem{Baz75} Ba\v zant, Z.P. (1975).``Theory of creep and shrinkage in
concrete structures: A pr\' ecis of recent developments", {\em Mechanics
Today}, ed. by S. Nemat-Nasser (Am. Acad. Mech.), Pergamon Press 1975, Vol.
2, pp. 1--93.

\bibitem{BazCed91} Ba\v zant, Z.P., and Cedolin, L. (1991). {\em Stability of
Structures: Elastic, Inelastic, Fracture and Damage Theories}, Oxford
University Press, New York; 2nd. ed. Dover Publ. 2003; 3rd ed. World
Scientific Publishing, Singapore--New Jersey--London 2010.

\bibitem{BazMos73} Ba\v zant, Z.P., and Moschovidis, Z. (1973). ``Surface
diffusion theory for the drying creep effect in Portland cement paste and
concrete" {\em J. Am. Ceramic Soc.}, 56, 235--241.

\bibitem{Bon-Pel10} Bonnaud, P.A., Coasne, B., and Pellenq, R.J.-M.
(2010). ``Molecular simulation of water confined in nanoporous silica."
{\em J. Phys.: Condens. Matter}, 284110 (15pp). 

\bibitem{Bro-Pel11} Brochard, L., Vandamme, M. and Pellenq, R.J.-M. (2011).
Competitive adsorption of carbon dioxide and methane in coal." Privately
communicated manuscript; submitted to PNAS

\bibitem{Bro-Pel11a} Brochard, L., Vandamme, M. and Pellenq, R.J.-M. (2011).
Poromechanics of nanoporous media." Privately communicated manuscript;
submitted to {\em J. of the Mech. and Phys. of Solids}.

\bibitem{Bru43} Brunauer, S. (1943). {\em The adsorption of gases and
vapors.} Princeton University Press, Princeton, NJ, 1943 (p. 398).

\bibitem{Coh38} Cohan, L. H. (1938) ``Sorption hysteresis and the vapor
pressure of concave surfaces", {\em J. Am. Chem. Soc.} 60, 430-435.

\bibitem{BET38} Brunauer, S., Emmett, P.T., and Teller, E. (1838).
``Adsorption of gases in multi-molecular layers." {\em J. Amer. Chemical
Soc.} 60, 309--319.

\bibitem{gelb1999} Gelb, L. D., Gubbins, K. E., Radhakrishnan, R. and
Sliwinska-Bartkowiak, M. (1999) "Phase separation in confined systems",
Rep. Prog. Phys. 62, 1573-1659.

\bibitem{Pel-Lev02} Pellenq, R. J.-M.  and Levitz, P. E. (2002) "Capillary
condensation in a disordered mesoporous medium: a grand canonical Monte
Carlo study", {\em Molecular Physics}, 100, 2059-2077.

\bibitem{Coa-Pel07} B. Coasne, A. Galarneua, F. Di Renzo, and R. J.-M.
Pellenq (2007) "Effect of morphological defects on gas adsorption in nano
porous silicas" {\em J. Phys. Chem. C}, 111, 15759-15770.

\bibitem{Coa-Pel09} B. Coasne, A. Galarneau, F. Di Renzo, and R. J.-M.
Pellenq (2009). ``Intrusion and retraction of fluids in nanopores: effect
of morphological heterogeneity", {\em J. Phys. Chem. C} 113, 1953-1962.

\bibitem{Coa-Pel08} B. Coasne, A. Galarneau, F. Di Renzo, and R. J.-M.
Pellenq (2008). ``Molecular simulation of adsorption and intrusion in
nanopores." {\em Adsorption} (2008) 14: 215–-221.

\bibitem{Coa-Pel08b} B. Coasne, F. Di Renzo, A. Galarneau and R.J.-M.
Pellenq (2008). ``Adsorption of simple fluid on silica surface and
nanopore: effect of surface chemistry and pore shape." {\em Langmuir} 24,
7285--7293

\bibitem{Der40} Derjaguin, B.V. (1940). ``On the repulsive forces between
charged colloid particles and the theory of slow caogulation and stability
of lyophole sols, {\em Trans. of the Faraday Society} 36, 203, 730.

\bibitem{EspFra06} Espinosa, R.M., and Franke, L. (2006). ``Influence of
the age and drying process on pore structure and sorption isotherms of
hardened cement paste." {\em Cement and Concrete Research} 36 (2006)
1969-–1984 (Figs. 2, 6, 16).

\bibitem{FelSer64} Feldman, R.F., and Sereda, P.J. (1964). ``Sorption of
water on compacts of bottle hydrated cement. I: The sorption and
length-change isotherms." {\em J. Appl. Chem.} 14, p.87.

\bibitem{FelSer68} Feldman, R.F., and Sereda, P.J. (1968). ``A model for
hydrated Portland cement paste as deduced from sorption-length change and
mechanical properties." Materials and Structures 1 (6) (1968) 509--520.

\bibitem{FreSmi96} Frenkel, D., and Smit, B. (1996). {\em Understanding
Molecular Simulations.} Academic Press, Sand Diego--New York.

\bibitem{Jen00} Jennings, H.M. (2000). ``A model for the microstructure of
calcium silicate hydrate in cement paste." {Cement and Concrete Research}
30, 101--116.

\bibitem{Jen10} Jennings, H.M. (2010). ``Pores and viscoelastic properties
of cement paste." submitted to an Elsevier Science journal.

\bibitem{Jen-Sch08} Jennings, H.M., Bullard, J.W., Thomas, J.J., Andrade,
J.E., Chen, J.J., and Scherer, G.W. (2008). ``Characterization and modeling
of pores and surfaces in cement paste: Correlations to processing and
properties." {\em J. of the Advanced Concrete Technology} 6 (1), 5--29.

\bibitem{Jon-Wen05} J\" onson, B., Nonat, A., Labbez, C., Cabane, B., and
Wennerstr\" om, H. (2005). ``Controlling the cohesion of cement paste."
{\em Langmuir} 21, 9211-9221.

\bibitem{JonWen04} J\" onson, B., Wennerstr\" om, H. Nonat, A., and Cabane,
B. (2004). ``Onset of cohesion in cement paste." {\em Langmuir} 20,
6702--6709.

\bibitem{Mal-Mur09} Malani, A., Ayappa, K.G., and Murad, S. (2009).
``Influence of hydrophillic surface specificity on the structural
properties of confined water." {\em J. Phys. Chem.} 113, 13825--13839.

\bibitem{Pel-Ulm09} Pellenq, R. J.-M., Kushima, A., Shashavari, R., Van
Vliet, K.J., Buehler, M.J., Yip, S. and Ulm, F.-J. (2010). ``A realistic
molecular model of cement hydrates." {\em Proc. Nat. Academy of Sciences}
106 (38), 16102--17107.

\bibitem{PowBro46} Powers, T.C., and T.L. Brownyard (1946), ``Studies of
the Physical Properties of Hardened Portland Cement Paste. Part 2. Studies
of Water Fixation". {\em J. of the Am. Concrete Institute} 18(3), 249--336.

\bibitem{Pow66} Powers, T.C. (1966). ``Some observations  on the
interpretation of creep data." {\em Bulletin RILEM (Paris)}, p. 381.

\bibitem{Rar-Jen95} Rarick, R.L., Bhatty, J.W. and Jennings, H.M. (1995).
``Surface area measurement using gas sorption: Application to cement
paste." {\em Material Science of Concrete IV}, editied by J. Skalny and S.
Mindess, Am. Ceramic Soc., Chapter 1, pp. 1--41.

\bibitem{Sch99} Scherer, G.W. (1999). ``Structure and properties of gels."
{\em Cement and Concrete Research} 29 (1999), 1149--1157.

\bibitem{Smi-Whi06} Smith, D.E., Wang, Y., Chaturvedi, A., Whitley, H.D.
(2006). ``Molecular simulations of the pressure, temperature, and chemical
potential dependencies of clay swelling." {\em J. Phys. Chem. B} 110,
20046--20054.

\bibitem{Tho-Jen08} Thomas, J.J., Allen, A.J., and Jennings, H.M. (2008).
``Structural changes to the calcium hydrate gel phase  of hydrated cement
with age, drying and resaturation." {\em J. of the Am. Ceramic Soc.} 91
(10), 3362--3369.


\bibitem{BazAsg-76} Ba\v zant, Z.P., Asghari, A. A., and Schmidt, J.
(1976). ``Experimental study of creep of hardened cement paste at variable
water content." {\em Materials and Structures} (RILEM, Paris),9, 279--190.

\bibitem{BazRaf82} Ba\v zant, Z.P., and Raftshol, W. J. (1982). ``Effect
of cracking in drying and shrinkage specimens." {\em Cement and Concrete
Research}, 12, 209--226; Disc. 797--798.


\bibitem{baroghel2007}  Baroghel-Bouny, V. (2007)
``Water vapour sorption experiments on hardened cementitious materials
Part I: Essential tool for analysis of hygral behaviour
and its relation to pore structure''
{\em Cement and Concrete Research} 37, 414Ð437.

\bibitem{romero2011} Romero-Vargas Castrill\'on, S., Giovambattista, N.,
Aksay, I. A., Debenedetti, P. G. (2011)  ``Structure and Energetics of Thin
Film Water," {\em J. Phys. Chem. C }, 115, 4624Ð4635.

\bibitem{suresh2000} Suresh, S.J., and Naik, V. M. (2000).``Hydrogen bond
thermodynamic properties of water from dielectric constant data". {\em
Journal of Chemical Physics}, 113, 9727?9732.


\bibitem{bazant1972} Ba\v zant, Z.P., and Najjar, L. J. (1972).
``Nonlinear water diffusion in nonsaturated concrete." {\em Materials and
Structures} (RILEM, Paris), 5, 3--20 {\small (reprinted in {\em Fifty Years
of Evolution of Science and Technology of Building Materials and
Structures}, Ed. by F.H. Wittmann, RILEM, Aedificatio Publishers, Freiburg,
Germany 1997, 435--456)}.

\bibitem{BazKap96}  Ba\v zant, Z.P., and Kaplan, M.F. (1996). {\em
Concrete at High Temperatures: Material Properties and Mathematical
Models}, Longman (Addison-Wesley), London (monograph and reference volume,
412 + xii pp.) (2nd printing Pearson Education, Edinburgh, 2002).



\bibitem{CerMed98} Cerofolini, G.F., Meda, L. (1998). ``A theory of
multilayer adsorption on rough surfaces in terms of clustering and melting
BET piles." {\em Surface Science} 416, 402--432.

\bibitem{CerMed98b} Cerofolini, G.F., Meda, L. (1998). ``Clustering and
melting in multilayer equilibrium adsorption." {\em J. of Colloid and
Interface Science} 202, 104--123.

\bibitem{Nik96} Nikitas, P. (1996). ``A simple statistical mechanical
approach for studying multilayer adsorption: Extensions of the BET
adsorption isotherm." {\em J. Phys. Chem.} 100, 15247--15254.

\bibitem{seri-levy1993} Seri-Levy, A. and Avnir, D. (1993) ``The
Brunauer-Emmett-Teller Equation and the effects of lateral interactions: A
simulation study", {\em Langmuir} 9, 2523-2529.

\bibitem{liu1993} Liu, H., Zhang, L., and Seaton, N. A. (1993) ``Analysis
of sorption hysteresis in mesoporous solids using a pore network model",
{\em Journal of Colloid and Interface Science} 156, 285-293.

\bibitem{hirth} 
Hirth, J. P. and Lothe, J., {\em Theory of Dislocations}, (Kreiger, 1992).


\bibitem{guggenheim} Guggenheim, E. A. {\em Mixtures} (Oxford University
Press, London, 1952).

\bibitem{CahHil58} Cahn, J. W.  and Hilliard, J. E.  (1958) ``Free energy
of a nonuniform system. I. Interfacial energy," {\em J. Chem. Phys.} 28,
258.

\bibitem{Van-Cou10} Vandamme, M., Brochard, L., Lecampion, B. and  Coussy,
O. (2010) ``Adsorption and strain: The CO2-induced swelling of coal," {\em
J. of the Mechanics and Physics of Solids} 58, 1489-1505.

\bibitem{thornton2003} Thornton, K., \AA gren, J. and Voorhees, P. W. (2003).
``Modelling the evolution of phase boundaries in solids at the
meso- and nano-scales",
{\em Acta Materialia} 51, 5675Ð5710.

\bibitem{baluffi}
Baluffi, R., Allen, S. A. and  Carter, W. C (2005) {\em Kinetics of
Materials} (Wiley, New York, 2005).

\bibitem{MZB_notes}, Bazant, Martin Z. (2011) {\em 10.626 Electrochemical
Energy Systems} (Massachusetts Institute of Technology: MIT
OpenCourseWare), http://ocw.mit.edu. License: Creative Commons BY-NC-SA.

\bibitem{meethong2007a} Meethong, N.,   Huang, H. Y. S., . Carter, W. C.,
and Chiang, Y. M.  (2007) ``Size-Dependent Lithium Miscibility Gap in
Nanoscale Li$_1-x$FePO$_4$", {\em Electrochemical and Solid-State Letters}
10, A134-A138.

\bibitem{meethong2007b}
Meethong, N.,   Huang, H. Y. S., Speakman, S. A. Carter, W. C., and Chiang,
Y. M.  (2007) ``Strain Accommodation during Phase Transformations in
Olivine?Based Cathodes as a Materials Selection Criterion for High?Power
Rechargeable Batteries", {\em Advanced Functional Materials } 17, 1115.

\bibitem{wagemaker2011}  Wagemaker, M., Singh, D. P.,  Borghols, W. J. H.,
Lafont, U.,  Haverkate, L., Peterson, V. K. and Mulder, F. M.  (2011)
``Dynamic Solubility Limits in Nanosized Olivine LiFePO$_4$", {\em J. Am.
Chem. Soc.}, 133, 10222Ð10228.

\bibitem{harris2010} Harris, S. J., Timmons, A., Baker, D. R., and Monroe,
C. (2010) ``Direct {\em in situ} measurements of Li transport in Li-ion
battery negative electrodes", {\em Chemical Physics Letters} 485, 265-274.

\bibitem{dreyer2010} Dreyer, W., Jamnik, J., Guhlke, C., Huth, R., Mo\v
skon, J., and Gaber\v s\v cek, M. (2010). ``The thermodynamic origin of
hysteresis in insertion batteries." {\em Nature Materials} 9, 448--451.

\bibitem{dreyer2011} Dreyer, D., Guhlke, C., and Huth, R. (2011) ``The
behavior of a many-particle electrode in a lithium-ion battery",
 {\em Physica D} 240, 1008Ð1019

\bibitem{bai2011} Bai, P., Cogswell, D. A., and Bazant, M. Z. (2011)
``Suppression of phase separation in LiFePO$_4$ nanoparticles during
battery discharge", preprint,  arXiv:1108.2326v1 [cond-mat.mtrl-sci]

\bibitem{singh2008} Singh, G. K. , Ceder, G.  and Bazant, M. Z.  (2008)
``Intercalation dynamics in rechargeable battery materials: General theory
and phase transformation waves in LiFePO$_4$,"  {\em Electrochimica Acta}
53, 7599.

\bibitem{burch2009} Burch, D. and Bazant, M. Z.  (2009) ``Size dependent
spinodal and miscibility gaps in nanoparticles," {\em Nano Letters} 9,
3795.

\bibitem{cogswell2011} Cogswell, D. A. and Bazant, M. Z. (2011)
``Coherency strain and the kinetics of phase separation in LiFePO$_4$",
preprint.

\bibitem{ferguson} Ferguson, T. R. and Bazant, M. Z., ``Phase
transformation dynamics in porous electrodes," in preparation. (Preliminary
results were presented the Electrochemical Society Meetings in Montreal in
May and in Boston in October, 2011.)

\bibitem{cahn1961} Cahn, J. W. (1961)  ``On spinodal decomposition", {\em
Acta Metallurgica}, 9, 795Ð801.

\bibitem{pahdi1997} Padhi, A. K., Nanjundaswamy, K. S. and Goodenough, J.
B. (1997) ``Phospho-olivines as positive-electrode materials for
rechargeable lithium batteries", {\em J. Electrochem. Soc.} 144, 1188-1194.

\bibitem{kang2009} Kang, B. and Ceder, G. (2009) ``Battery materials for
ultrafast charging and discharging",  {\em Nature} 458, 190-193.

\bibitem{malik2010}  Malik, R., Burch, D., Bazant, M. and Ceder, G. (2010)
``Particle Size Dependence of the Ionic Diffusivity", {\em Nano Letters}10,
4123-4127.

\bibitem{malik2011} Malik, R., Zhou, F. and Ceder, G. (2011) ``Kinetics of
non-equilibrium lithium incorporation in LiFePO$_4$",  {\em Nature
Materials}, 10 (8), 587-590.

\bibitem{pellenq1997}  Pellenq, R. J.-M.,  Caillol, J. M.,  and Delville,
A. (1997) "Electrostatic Attraction between Two Charged Surfaces: A (N,V,T)
Monte Carlo Simulation", {\em J. Phys. Chem. B} 101, 8584-8594.

\bibitem{RTIL2011} Bazant, M. Z., Storey, B. D., and Kornyshev, A. A.
(2011) "Double layer in ionic liquids: Overscreening versus crowding", {\em
Physical Review Letters}, 106, 046102.

\bibitem{nauman1989} Nauman, E. B. and Balsara, N. P. (1989) ``Phase
equilibria and the Landau-Ginzburg functional", {\em Fluid Phase
Equilibria}, 45, 229-250.

\bibitem{nauman1991} Nauman, E. B., Rousar, I. and Dutia, A. (1991) ``On
the ultimate fineness of a dispersion", {\em Chemical Engineering
Communications},  105, pp. 61-75.



\bibitem{degennes1985} de Gennes, P. G. (1985) ``Wetting: statics and
dynamics", Rev. Mod. Phys. 57, 827-863.

\bibitem{israelachvilli} Israelachvili, J. {\it Intermolecular and Surface
Forces} (Academic Press, New York, 1992).

\bibitem{rowlinson} Rowlinson, J. S.  and Widom, B. {\it Molecular Theory
of Capillarity} (Clarendon Press, Oxford, 1984).

\bibitem{cahn1977} Cahn, J. W. (1977) ``Critical point wetting", J. Chem.
Phys. 66, 3667 (1977).

\bibitem{gouin2008} Gouin H. and Gavrilyuk, S. (2008) ``Dynamics of liquid
nanofilms", Int. J. Eng. Sci. 46, 1195-1202.

\bibitem{gouin2009} Gouin H. (2009) ``Liquid nanofilms: A mechanical model
for the disjoining pressure ", Int. J. Eng. Sci. 47, 691-699.

\bibitem{vdw1893} van der Waals, J. D. (1893). The thermodynamic theory of
capillarity under the hypothesis of a continuous variation of density.
(Translation by J. S. Rowlinson (1979)) Journal of Statistical Physics, 20,
197 (Original version: Zeitscherift fuer Physikalische Chemie,
Stoechiometrie und Verwandtschaftslehre, 13, 657).

\bibitem{widom1999} Widom, B. (1999) ``What do we know that van der Waals
did not know?", Physica A, {\bf 263}, 500-515.

\bibitem{naumann2001} Naumann, E. B. and He, D. Q. (2001) ``Nonlinear
diffusion and phase separation", {\em Chem. Eng. Science} 56, 1999-2018.

  }
\end{thebibliography}
\end{document}